\documentclass[twocolumn, tighten, times]{aastex61}
\usepackage{graphicx}
\usepackage{epstopdf}

\usepackage{threeparttable}
\usepackage{sidecap}
\sidecaptionvpos{figure}{c}
\usepackage{floatrow}
\shorttitle{Asymmetric star formation triggered by gas inflow in a barred lenticular galaxy PGC 34107}
\shortauthors{Lu et al. }
\begin{document}

\title{Asymmetric star formation triggered by gas inflow in a barred lenticular galaxy PGC 34107}

\correspondingauthor{Qiusheng Gu}
\email{qsgu@nju.edu.cn}

\author{Shiying Lu}
\affil{School of Astronomy and Space Science, Nanjing University, Nanjing, Jiangsu 210093, China }
\affil{Key Laboratory of Modern Astronomy and Astrophysics (Nanjing University), Ministry of Education, Nanjing 210093, China}

\author{Qiusheng Gu}
\affil{School of Astronomy and Space Science, Nanjing University, Nanjing, Jiangsu 210093, China }
\affil{Key Laboratory of Modern Astronomy and Astrophysics (Nanjing University), Ministry of Education, Nanjing 210093, China}

\author{Xue Ge}
\affil{School of Physics and Electronic Engineering, Jiangsu Second Normal University, Nanjing, Jiangsu 211200, China}

\author{Luis C. Ho}
\affil{Kavli Institute for Astronomy and Astrophysics, Peking University, Beijing 100871, China}
\affil{Department of Astronomy, School of Physics, Peking University, Beijing 100871, China}

\author{Yulong Gao}
\affil{School of Astronomy and Space Science, Nanjing University, Nanjing, Jiangsu 210093, China }
\affil{Key Laboratory of Modern Astronomy and Astrophysics (Nanjing University), Ministry of Education, Nanjing 210093, China}

\author{Zhengyi Chen}
\affil{School of Astronomy and Space Science, Nanjing University, Nanjing, Jiangsu 210093, China }
\affil{Key Laboratory of Modern Astronomy and Astrophysics (Nanjing University), Ministry of Education, Nanjing 210093, China}

\author{Ke Xu}
\affil{School of Astronomy and Space Science, Nanjing University, Nanjing, Jiangsu 210093, China }
\affil{Key Laboratory of Modern Astronomy and Astrophysics (Nanjing University), Ministry of Education, Nanjing 210093, China}

\author{Zhi-Yu Zhang}
\affil{School of Astronomy and Space Science, Nanjing University, Nanjing, Jiangsu 210093, China }
\affil{Key Laboratory of Modern Astronomy and Astrophysics (Nanjing University), Ministry of Education, Nanjing 210093, China}

\author{Yong Shi}
\affil{School of Astronomy and Space Science, Nanjing University, Nanjing, Jiangsu 210093, China }
\affil{Key Laboratory of Modern Astronomy and Astrophysics (Nanjing University), Ministry of Education, Nanjing 210093, China}

\author{Qirong Yuan}
\affil{Department of Physics and Institute of Theoretical Physics, Nanjing Normal University, Nanjing 210023, China }

\author{Min Bao}
\affil{Department of Physics and Institute of Theoretical Physics, Nanjing Normal University, Nanjing 210023, China }
\affil{School of Astronomy and Space Science, Nanjing University, Nanjing, Jiangsu 210093, China }
\affil{Key Laboratory of Modern Astronomy and Astrophysics (Nanjing University), Ministry of Education, Nanjing 210093, China}

\begin{abstract}
Comparing to the inactive and gas-poor normal lenticular galaxies (S0s) in the local universe, we study a barred star-forming S0 galaxy, PGC 34107, which has been observed by the Centro Astron\'{o}mico Hispano Alem\'{a}n (CAHA) 3.5-m telescope and the Northern Extended Millimeter Array (NOEMA). The spatially resolved ionized gas and molecular gas traced by $^{12}$CO(1-0), hereafter CO(1-0), show the similar distribution and kinematics to the stellar component with an off-center star-forming region, $\sim$380 pc away from the center. The resolved kinematics of molecular CO(1-0) emission reveals that there is a blueshifted (redshifted) velocity component on the receding (approaching) side of the galaxy along the stellar bar. This might provide a plausible evidence of non-circular motion, such as the bar-induced molecular gas inflow. The velocity of molecular gas inflow decreases with approaching towards the peak of the off-center star formation in the north, which might be associated with the inner Lindblad resonance (ILR). In addition to CO(1-0), we also detect the isotopic line of $^{13}$CO(1-0). Most $\rm H\alpha$, CO(1-0) and $^{13}$CO(1-0) emissions are concentrated on this northern star-forming region. We find that PGC 34107 follows the local stellar mass-metallicity relation, star-forming main sequence, and the Kennicutt-Schmidt law. The resolved and integrated molecular gas main sequence suggest that there is a higher gas fraction in the galaxy central region, which supports a scenario that the bar-induced gas reservoir provides the raw material, and subsequently triggers the central star formation.

\end{abstract}


\section{Introduction}
\label{sec1:introduction}
Lenticular galaxies (S0s) are traditionally considered to be at an intermediate position between elliptical and spiral galaxies (\citealt{Hubble+1936}). S0s not only display the discy structure seen in spirals, but also contain the red bulge with old stellar populations typically found in ellipticals. Thus, normal S0s, as a class of early-type galaxies (ETGs), are usually gas-poor and inactive (e.g., \citealt{Bregman+1992,Caldwell+1993,Blanton+09}). PGC 34107 shows an early-type morphology, and active star formation in its center (e.g. \citealt{Contini+1998, Xiao+16, Zhou+20}). Based on the Canada-France-Hawaii Telescope (CFHT) $r$-band image, \cite{Zhou+20} analyzed the structure of PGC 34107 from the 1-D profile and 2-D multi-component decomposition, and found that PGC 34107 is a barred S0 galaxy with a disk-like pseudo-bulge (S{\'e}rsic index $n$=1.29), which is consistent with other star-forming S0s as found by \cite{Xiao+16} and \cite{Mishra+17}. At the meantime, \cite{Alatalo+13} detected a limited amount of molecular gas in PGC 34107. Therefore, PGC 34107 is a special and promising candidate of S0s with signs of star formation and cool gas. However, we can not completely rule it out as a possibly misclassified normal spiral galaxy.

A commonly accepted scenario of the formation of S0s is that they are the end product of the evolution of spiral galaxies stripped out of gas (\citealt{Spitzer+Baade+1951, Dressler+1980, D'Onofrio+15}). The increasing fraction of S0s with decreasing redshift (e.g., \citealt{Postman+05, Desai+07}) and the observation of gas stripping of spirals (e.g., \citealt{Vollmer+09,Sivanandam+10}) can support the transformation of S0s from spiral galaxies. In S0s, their globular cluster specific frequency (e.g., \citealt{Barr+07}), mass-metallicity trends (e.g., \citealt{Prochaska+Chamberlain+11}) and stellar kinematics inferred from planetary nebulae (e.g., \citealt{Cortesi+11}) are consistent with those observed and derived in spiral galaxies, which is also in favor of the transformation from spiral galaxies. During the transformation, the environment may play an essential role of stripping out of gas from spiral arms (e.g., \citealt{Postman+Geller+1984, Dressler+1980}). Comparing to the field galaxies, the higher frequency of normal S0s in groups/clusters hints that S0s prefer to form in dense environment by interactions, such as  galactic winds (e.g., \citealt{Ho+14}), ram pressure strapping (\citealt{Gunn+Gott+1972}), strong tidal interactions and mergers (e.g., \citealt{Barnes+Hernquist+1992,Mazzei+14a, Mazzei+14b}), and galaxy harassment (e.g., \citealt{Moore+1996}). 
Based on the Mapping Nearby Galaxies at APO Data Release 15 (MaNGA-DR15) survey, \cite{Fraser-McKelvie+18} and \cite{Dominguez+Sanchez+20} supplemented that S0s with different masses may be formed through different physical processes. They proposed that the spiral-to-S0 fading scenario likely formed less massive ($\rm <10^{10} M_{\odot}$) S0s by the bulge rejuvenation (\citealt{Fraser-McKelvie+18}) or in-situ star formation (\citealt{Dominguez+Sanchez+20}), while massive ($\rm >10^{10} M_{\odot}$) S0s transformed via morphological/inside-out quenching (\citealt{Fraser-McKelvie+18}) or gas-rich mergers (\citealt{Dominguez+Sanchez+20}).

However, there are arguments against the spiral-to-S0 fading scenario. If S0s are stripped spirals whose star formation has been quenched, we expect the bulges of S0s are connected to those of spirals. While \cite{Gao+18} performed multi-component decomposition of S0s by using high-quality optical images from the Carnegie-Irvine Galaxy Survey, and found that the bulges of late-type spirals and S0s were intrinsically different, which implied that spirals were not the plausible progenitors of S0s. Similarly, \cite{Burstein+05} pointed out if S0s were gas-stripped spiral galaxies, the absolute K magnitudes of S0s would be expected to be $\sim$0.75 mag less luminous than early-type spirals in models, but they didn't find this difference.
Meanwhile, \cite{Silchenko+12} found the higher metallicity in the outer disks of nearby S0s, and \cite{Holden+09} suggested that there is no evolution in the overall distribution of bulge-to-disk ratios for cluster early-type galaxies from z$\sim$0 to z$\sim$1. All results suggest that there may exist other physical processes that can form S0s (e.g., \citealt{Kormendy+Kennicutt+04, Barway+09, van+den+Bergh+09}).

Although S0s are often thought to be inactive, the nuclear star formation in S0s has been found by some studies (\citealt{Schawinski+07, Kaviraj+07, Xiao+16, Fraser-McKelvie+18}). \cite{Welch+Sage+03} and \cite{Welch+10} detected the cool neural and/or molecular gas in more than 50 percents of normal S0s. This reservoir of gas accreted from environment (\citealt{Dressler+13,Silchenko+19}) provides the raw material, then galaxy interactions or gas-rich mergers can trigger the star formation of S0s (\citealt{Thilker+10, Davis+15, Xiao+16, Ge+20}). 
As the expansion of the universe and the virialization of galaxy clusters, major mergers become less common and the internal secular processes gradually become dominant with the cosmic time (\citealt{Kormendy+Kennicutt+04}). 
The relevant secular processes can generally shape many distinct substructures of S0s in low density environment, such as bars, ovals and lenses (e.g., \citealt{Laurikainen+05,Laurikainen+09}). Bars have been proven to trigger the secular evolution of disk galaxies by theory and numerical simulations (\citealt{Athanassoula+1992,Sellwood+1993,Piner+1995,Knapen+00,Athanassoula+03}). 
More spatially resolved observations of the cool molecular gas can help reveal the kinematics of the gaseous bar and infer the formation of S0s.

\begin{figure}
\centering
\includegraphics[width=0.85\textwidth]{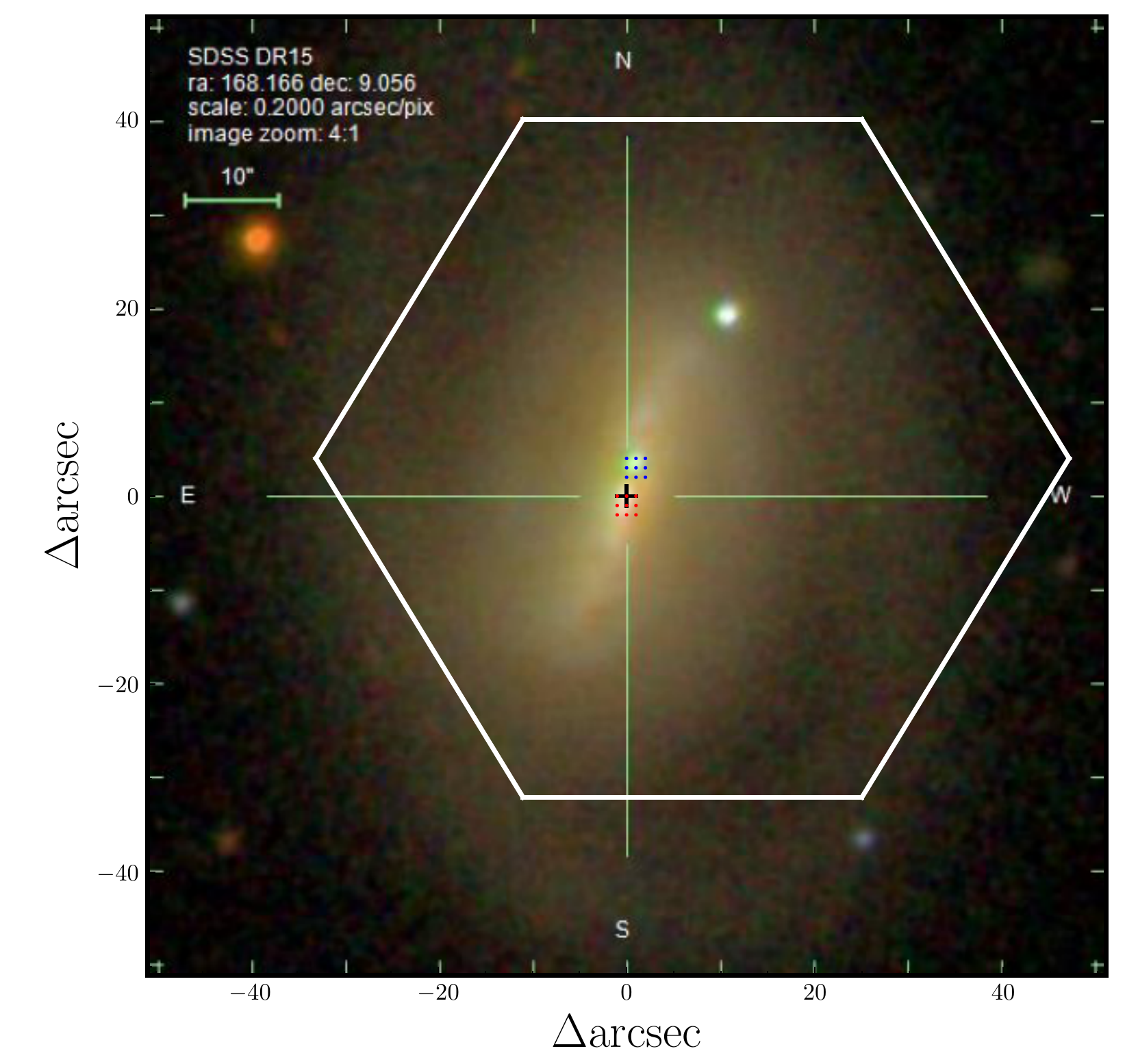}
\caption{The false-color ($g$-, $r$- and $i$-bands) image of PGC 34107. The white hexagon shows the CAHA bundle allocation, and the black cross presents the optical center of galaxy. The blue and red dots represent the corresponding brightness spaxels used to derived the local spectra in the south and north (see section~\ref{sect2.1:CAHA Observation} for details). }
\label{fig1}
\end{figure}

In this work, PGC 34107 is a nearby (z=0.00471) barred S0 galaxy at  a distance of $\sim$ 20.2 Mpc, which corresponds to $1\arcsec$$\sim$97 pc. PGC 34107 is one of star-forming S0 galaxies (SFS0s) from \cite{Xiao+16}. Based on the cross-matching catalog between Sloan Digital Sky Survey Data Release 7 (SDSS DR7) and the Third Reference Catalog of Bright Galaxies (RC3), \cite{Xiao+16} found that SFS0s with lower S\'{e}rsic indices and stellar masses, mainly live in a sparse environment. It is what kind of secular processes that triggers the nuclear star formation of S0s in the sparse environment, resulting in the difference between SFS0s and normal S0s. Unlike SFS0s triggered by gas-rich minor merger, PGC 26218 and PGC 38025, which have been studied by \cite{Ge+20} and \cite{Chen+21}, respectively, our target PGC 34107 has a stellar bar and two bright regions, which are also called double nuclei by \cite{Zhou+20}, where they presented long-slit spectroscopy along the major axis, and concluded that the double nuclei may be formed and evolved by secular processes driven by the stellar bar or the external accretion of gas. Here we present the spatially resolved optical observation with the Centro Astron\'{o}mico Hispano Alem\'{a}n (CAHA) 3.5-m telescope and millimeter observation with the NOrthern Extended Millimeter Array (NOEMA), respectively, which can help to reveal the kinematics of stellar component, ionized and molecular gas. Compared to previous works, what is important is that tentative signs of inflow along the bar traced by the molecular gas are found due to the high quality of NOEMA data, which provides a possible evidence of the star formation induced by the stellar bar in PGC 34107. Table~\ref{tab1} summaries the global parameters derived from different works. Figure~\ref{fig1} shows the SDSS $gri$ image of PGC 34107. The optical center of PGC 34107 is marked with a black cross. There are two obvious bright regions, i.e., northern blue and southern red regions, seen near its center, while the bright point in the northwest is a foreground star.

\begin{deluxetable*}{lcccc}
\tablecaption{The Global Properties of PGC 34107\label{tab1}}
\tablehead{
\colhead{} & \colhead{ This work}& \colhead{}&\multicolumn2c{Pervious works} \\
\cline{4-5} 
\colhead{Property}& \colhead{Value}&\colhead{} & \colhead{Value} & \colhead{Reference}
}
\startdata
R.A. (J2000)& --&  &168.165872 & SDSS\\
Decl. (J2000)& --& & +09.055836 & SDSS \\
Redshift & -- & & 0.00471& SDSS  \\
$\rm \log M_*$ ($\rm M_\odot $)& $9.60\pm 0.001$ & & 9.87$^{s}$/9.13& MPA-JHU DR7/\cite{Zhou+18}  \\
$\rm \log SFR$ ($\rm M_\odot yr^{-1}$) & $-0.48\pm 0.01$&  &-0.37$^{s}$/-0.56$^{s}$ & MPA-JHU DR7/\cite{Zhou+18}  \\
$\rm \log \Sigma_{SFR}$ ($\rm M_\odot yr^{-1} kpc^{-2}$) & $-0.51\pm 0.08$ & &$-0.33\pm 0.19^{s}$ & \cite{Davis+14} \\
$\rm \log M_{H_2}$ ($\rm M_\odot$) & $8.22\pm 0.13$ & & 8.28& \cite{Young+11} \\
						        &                           & & 8.36  &\cite{Alatalo+13}  \\
$\rm \log \Sigma_{H_2}$ ($\rm M_\odot \; pc^{-2}$) &$2.34\pm 0.13$ & & -- & -- \\
$\rm \log \Sigma_{gas}$ ($\rm M_\odot \; pc^{-2} $) & $ 2.52\pm 0.14$& & $ 2.31\pm 0.22$ &\cite{Davis+14}  \\
\enddata
\tablecomments{The superscript ``s'' indicates that parameters are recompiled by adopting a \cite{Salpeter+1955} IMF instead of an original \cite{Kroupa+01}. \cite{Kauffmann+03}, \cite{Brinchmann+04} and \cite{Tremonti+04} derived the catalogs of MPA-JHU DR7. In this work, the $\rm M_{H2}$ and $\rm \Sigma_{H_2}$ are derived within the CO region ($\rm \sim 0.63 kpc^2$) in Figure~\ref{fig4} at the luminosity distance ($D_{\rm L}$) of 20.2 Mpc. Note that the masses ($\rm M_{H2}$) of \cite{Young+11} and \cite{Alatalo+13} are converted by adopting a same $\rm \alpha_{CO}$ and $D_{\rm L}$ as this work. The $\Sigma_{\rm gas}$ is calculated within the same $\rm H\alpha$  and CO coverage, as shown in the solid black polygon of Figure~\ref{fig10}. }
\end{deluxetable*}
   
This paper is laid out as follows. Section \ref{sect2:CAHA} and \ref{sect3:NOEMA} show the observations and simple data analyses of the optical Integral Field Unit (IFU) and millimeter data, respectively. In section \ref{sect4:Discussion}, we present the discovery of gas inflow, gas distributions (i.e., $\rm H\alpha$, $^{12}$CO(1-0) and  $^{13}$CO(1-0)), the global mass-metallicity relation (MZR) and star-forming main sequence (SFMS), and Kennicutt-Schmidt (K-S) law and molecular gas main sequence (MGMS) by combining the 2D-spectroscopic observation with the millimeter observation, respectively. In section \ref{sect5:Summary}, we present our summary. Throughout this paper, we assume a flat $\Lambda$CDM cosmology with $\rm \Omega_{M} = 0.3$, $\rm \Omega_{\Lambda} = 0.7$, and $\rm H_{0} = 70\, \rm km\;s^{-1}~Mpc^{-1} $ and a \cite{Salpeter+1955} initial mass function (IMF).

\section{CAHA Optical Data}
\label{sect2:CAHA}

\begin{figure*}[t]
  \begin{center}
  \includegraphics[width=0.8\textwidth, angle=0 ]{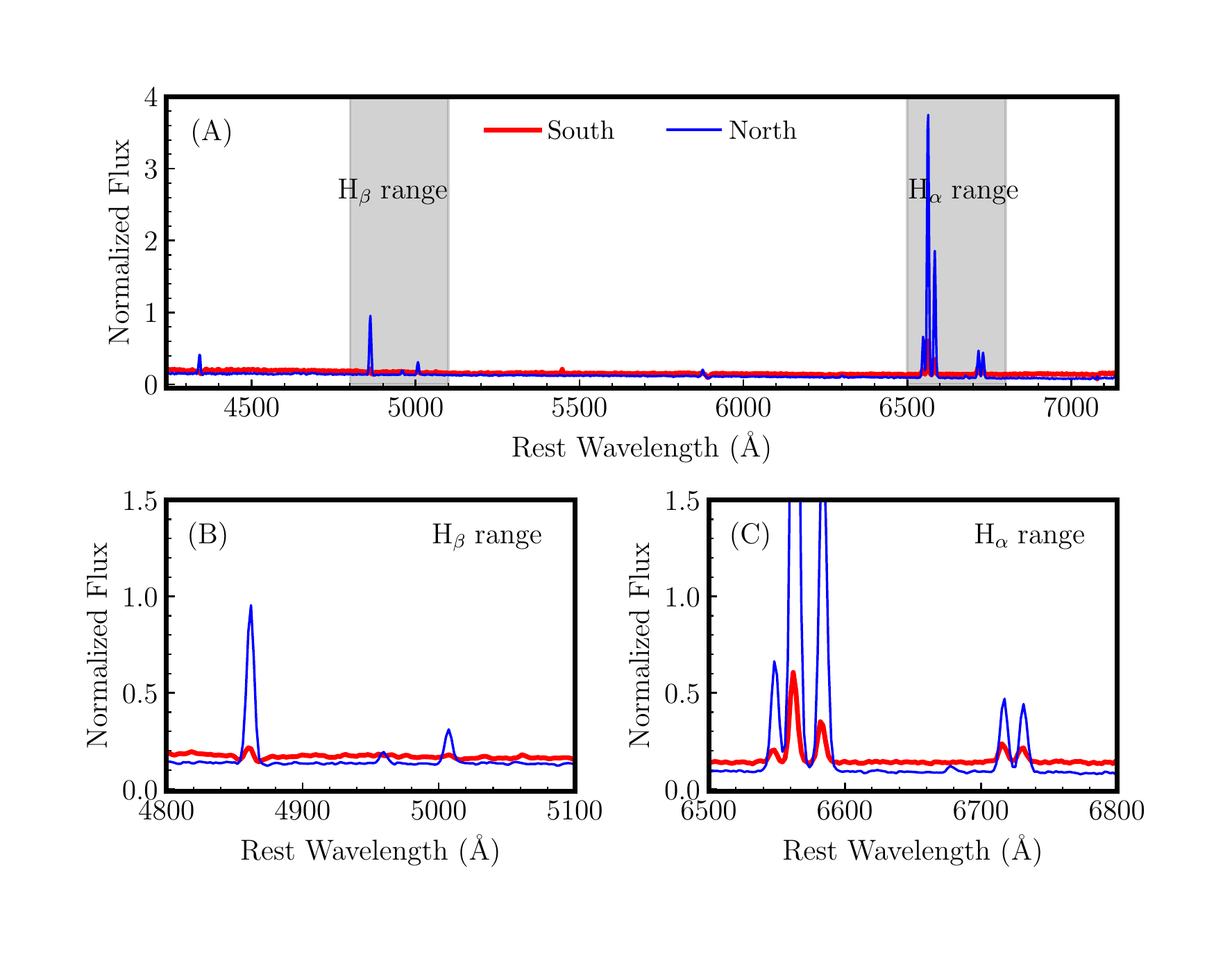}
  \end{center}
  \caption{
The residual spectrum in blue (red) between the average stacked spectrum of 9 spaxels in the north (south) and the average stacked spectrum of 12 spaxels around those 9 spaxels. 
The corresponding positions of the 9 northern (southern) spaxels are shown with blue (red) dots in the Figure~\ref{fig1}. The spectra are normalized to $\rm \sim 5100\AA$ flux of the central spaxel of PGC 34107, and the signal-to-noise (S/N) of each spaxel spectrum is larger than 10. The upper panel A presents the full spectra, while the zoom-in spectra in $\rm H\beta$ and $\rm H\alpha$ ranges are shown in the bottom panel B and C, respectively. There are only emission lines left in the residual spectra, which implies that they are two star-forming regions, and not star-forming nuclei.}
  \label{fig2}
\end{figure*}

\subsection{CAHA 2D-spectroscopic Observation}
\label{sect2.1:CAHA Observation}

\begin{figure*}[ht]
  \begin{center}
  \includegraphics[width=0.80\textwidth, angle=0 ]{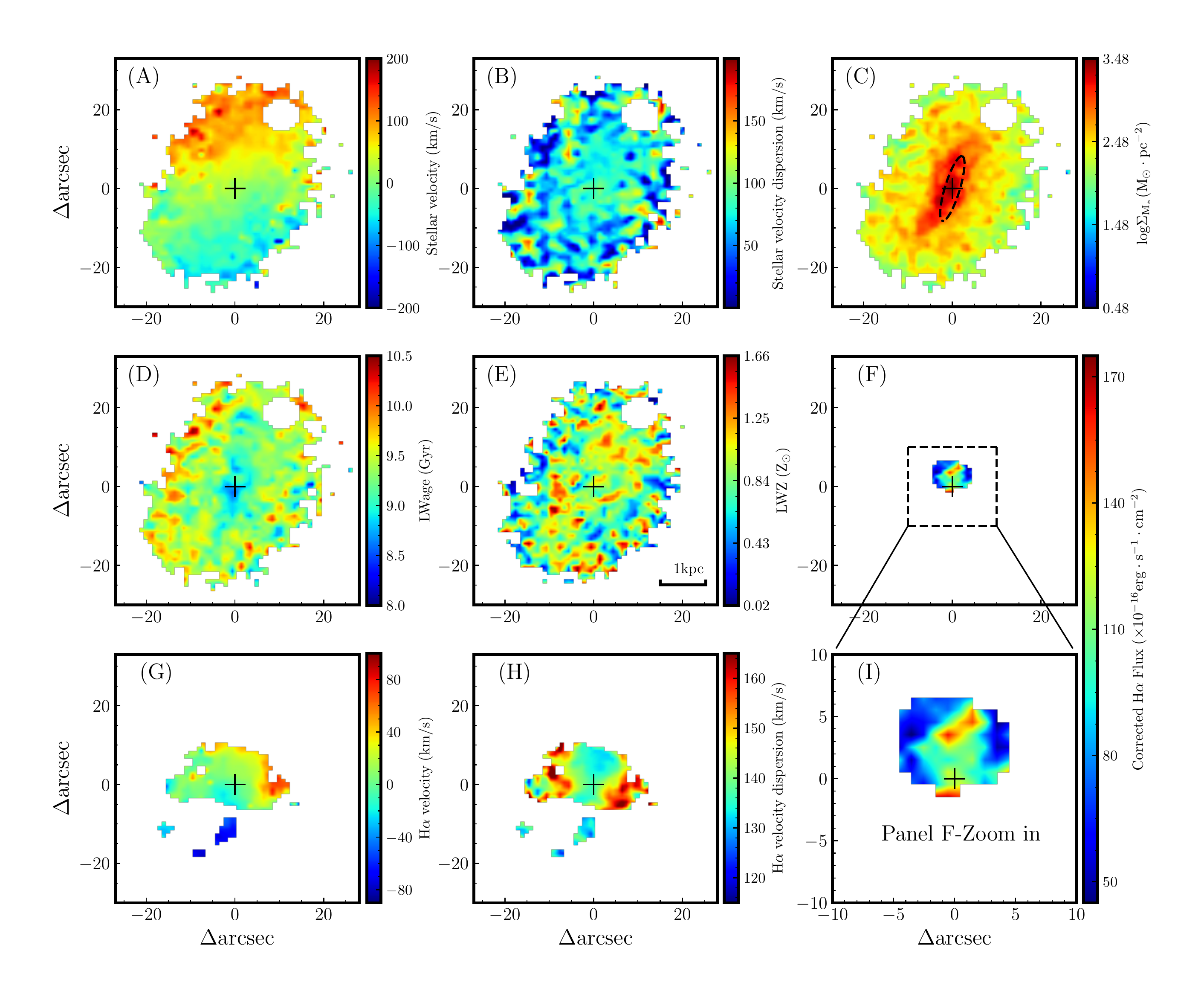}
  \end{center}
  \caption{Distributions of physical parameters color coded by (Panel A) the stellar velocity, (B) the stellar velocity dispersion,(C) the surface mass density ($\rm M_* \cdot pc^{-2}$), (D) the light-weighted age (LWage), (E) the light-weighted metallicity (LWZ),  (F) the corrected $\rm H\alpha$ flux, (G) the $\rm H\alpha$ velocity and (H) the $\rm H\alpha$ velocity dispersion, respectively. The zoom-in map of Panel F is shown in Panel I for a better view.  The black dashed ellipse in Panel C represents the position and length of the stellar bar, which is obtained by performing 2D multi-component decomposition of r-band SDSS image with GALFIT (\citealt{Peng+02, Peng+10}). The white circle in the northwest is corresponding to the position of the foreground star. The S/Ns of emission lines, including $\rm H\alpha$ and $\rm H\beta$, are imposed to be larger than 5 (Panels G and H: S/N$_{\rm H\alpha}>$5, Panels F and I: S/N$_{\rm H\alpha, H\beta}>$5). The map center marked with a black cross is set at 168.166 (RA), +09.056 (DEC). }
  \label{fig3}
\end{figure*}

We started a program to obtain Integral Field Spectroscopy (IFS) data of SFS0s in \cite{Xiao+16}, with the PMAS/PPAK configuration mounted on the CAHA 3.5-m telescope in the Calar Alto observatory.
The observations were performed during March 2016 and April 2017. The optical IFU spectroscopic observation of PGC 34107 was carried out on March 31 2017 with the low-resolution (V500) setup, covering a hexagonal field-of-view (FoV) of $78\arcsec \times73\arcsec$, and a wavelength of 3745-7500 $\rm \AA$ at a spectral resolution of 6 $\rm \AA$ (R$\sim$850).  After examining the internal vignetting effect of blue and red part of spectra from fibers around the FoV center, the wavelength range from 4240 to 7140$\rm \AA$ is adopted in this work (\citealt{Husemann+13}). In order to reach a filling factor of 100\% across the entire FoV, a three-pointing dithering scheme has been taken, thus the final 3D data cube is composed of 4221 spectra at a sampling of $1\arcsec$$\times$$1\arcsec$ per spaxel. The exposure time per pointing is 900 seconds. The atmospheric seeing is about 2\arcsec, but it is not a limiting factor of spatial resolution because the final spatial resolution of the CALIFA data is set by fiber size and the dither scheme together with the adopted image reconstruction algorithm (\citealt{Garcia-Benito+15}).

We used an upgraded python-based pipeline (\citealt{Garcia-Benito+15, Sanchez+16}) to reduce the PPAK IFU data. The data were reduced though a series of processes, including: 1) removal of electronic signatures and realignment of the frames; 2) spectral extraction, wavelength calibration and fiber transmission correction; 3) sky subtraction; 4) flux calibration; 5) spatial re-arranging and image reconstruction; 6) differential atmospheric refraction; and 7) absolute flux re-calibration. More detailed data reduction processes are described in \cite{Sanchez+12}, \cite{Husemann+13}, \cite{Garcia-Benito+15} and \cite{Sanchez+16}. Finally, the reprocessed spectral data are stored in a 3D data cube. In the ATLAS$^{\rm 3D}$ project, PGC 34107 had been observed for 2 hours by SAURON IFS mounted on the William Herschel Telescope on La Palma (\citealt{Cappellari+11}), which had a FoV of $34\times41$ arcsec$^2$, a spectral resolution of 4.2 $\rm \AA$ and a wavelength range of 4800-5380 $\rm \AA$. Compared to the SAURON observation, although our exposure time and spectral resolution is shorter and lower, we have a longer wavelength range, including the $\rm H\alpha$, $\rm [NII]\lambda \lambda6548, 6583$ and $\rm [SII] \lambda \lambda 6717, 6731$, which are important emission lines to understand the photoionization.

As shown in Figure~\ref{fig1}, there are two bright regions in the central region of PGC 34107. For each bright region (north or south), we can compare a spectrum derived from the central spaxels with a spectrum derived from the periphery. 
If the bright region is a nucleus, experiencing a mass concentration, the contribution from stellar continuum could be detected in the residual spectrum obtained by subtracting the outer bright region spectrum from the inner bright region. Conversely, if only the emission lines are left, it suggests that this bright region is only recent star formation events rather than nuclei. 
Therefore, we first stack the spectra of 9 spaxels in the northern and southern brightness peaks shown in Figure~\ref{fig1}, and take their average spectra to be the local spectrum. Next, we stack 12 spectra around the northern/southern brightness peaks, and take the average to represent the surrounding spectrum. After subtracting the surrounding spectrum from the local spectrum, the residual spectra in the north (blue) and south (red) are shown in Figure~\ref{fig2}. Both spectra have been normalized to the $\rm \sim 5100 \AA$ flux of the central spaxel of PGC 34107, and the position of central spaxel is shown on black cross in Figure~\ref{fig1}. As shown in Figure \ref{fig2}, the stellar continua are almost subtracted and only emission lines are left, which imply that both blue and red bright regions in PGC 34107 tend to be the star-forming regions rather than mass concentrations or nuclei.

\subsection{Full Spectral Fitting}
\label{sect2.2: Full Spectral Fitting}

For each spectrum in the datacube of PGC 34107,  we adopted a public code, the penalized pixel-fitting (pPXF),  to decompose  stellar and emission-line components via full spectral fitting (\citealt{Cappellari+Emsellem+04,Cappellari+17}). Following \cite{Ge+20}, for each spaxel we adopt the average flux and the standard deviation of flux covering the wavelength range 5075-5125 $\rm \AA$ as the signal and noise as well, because such range can avoid the contamination of emission and absorption lines. The signal-to-noise (S/N) ratio of the most spaxels of PGC 34107 within the effective radius ($<$Re) is more than 10, which is high enough so that no more information can be provided by using Voronoi binning method (\citealt{Cappellari+Copin+03}). So in this work, spectral fitting of each spaxel with S/N $>10$ is performed separately, without Voronoi binning.

With emission lines being masked out, the stellar component in the rest-frame spectrum is fitted with simple stellar population (SSP) templates of MILES (\citealt{Sanchez-Blazquez+06,Vazdekis+10}), assuming a \cite{Salpeter+1955} IMF and a \cite{Calzetti+00} dust extinction law. The SSP templates evenly distribute on the age-metallicity grid, with 25 ages ranging from 0.06 to 15.85 Gyr and 6 different metallicities ($\rm \log Z/Z_\odot$=-1.71, -1.31, -0.71, -0.4, 0.0, 0.22). Figure~\ref{fig3} shows the distributions of physical properties of stellar population derived by fitting, including the stellar velocity (Panel A), the stellar velocity dispersion (Panel B), the surface mass density (Panel C), the light-weighted age (Panel D) and the light-weighted metallicity (Panel E). The stellar velocity map (Panel A) shows a characteristic rotating disk-shape structure, 
and the rotation of the galaxy should be counter-clockwise on the sky in order for the spiral arms to be trailing.
The average velocity dispersion of stellar shown in Panel B is about $\sim$ 97 km $\rm s^{-1}$. Panel C shows a higher stellar mass surface density along the bar than the outskirt of galaxy. The position and length of bar is indicated by a black dashed ellipse in Panel C, which is derived by performing 2D multi-component decomposition of r-band SDSS image with GALFIT (\citealt{Peng+02,Peng+10}). Following \cite{Zhou+20}, we adopt three S{\'e}rsic models for the bar, bulge and disk of PGC 34107, respectively. The position angle ($\rm PA_{bar}$, in degrees east of north) and radius ($\rm R_{bar}$) of bar is -15.97$^{\circ}\pm$0.03$^{\circ}$ and 8.4$\arcsec \pm$0.03$\arcsec$, respectively, which is consistent with results ($\rm PA_{bar}$=-17.58$^{\circ}$ and $\rm R_{bar}$=10.22$\arcsec$) of \cite{Zhou+20}. The total stellar mass is estimated to be about $\rm 4.0\times 10^{9} M_\odot$, which is summarized in Table~\ref{tab1}. It is $\sim$0.27 dex lower than that from the MPA-JHU DR7 catalog, but $\sim$0.47 dex higher than that from \cite{Zhou+18}, which is derived by using WISE 3.4 $\rm \mu m$ luminosity. Note that stellar masses in the MPA-JHU DR7 catalog are based on the \cite{Kroupa+01} IMF, while we adopt the \cite{Salpeter+1955} IMF in this work. The stellar mass listed in Table~\ref{tab1} has been converted to the \cite{Salpeter+1955} IMF. Maps of the light-weighted age and metallicity are shown in Panels D and E, respectively. 
The distribution of metallicity is relatively uniform, while the central distribution of age is younger than that in the outskirts, which suggests that there is an active star formation in the central region of the galaxy. 
Note that when fewer available spaxels with the wavelength range from 3750 to 7500 $\rm \AA$ are used to fit the pPXF, the light-weighted age near the center is still younger than the outer region of the galaxy.

For the emission lines, we subtract the best-fit synthesized stellar continuum from the observed spectrum to get the pure emission-line spectrum. Subsequently, each emission line is fitted with one single Gaussian profile with IDL package MPFIT (\citealt{Markwardt+09}). Following \cite{Xiao+16}, $\rm H\beta$ and $\rm [OIII]\lambda\lambda 4959, 5007$ lines are fitted simultaneously, while $\rm H\alpha$ and $\rm [NII]\lambda \lambda6548,6583$ are fitted together. The S/N for each emission line is then estimated using the method introduced in \cite{Ly+14}, where the flux is determined by 
\begin{equation}
   \textit{\rm Flux} =\sum_{-2.5\sigma_{\rm G}}^{+2.5\sigma_{\rm G}}[f(\lambda - l_C)- \langle f \rangle] \times l',
\label{eq1}
\end{equation}
and the noise is estimated by 
\begin{equation}
   \textit{\rm Noise} =\sigma(f) \times l' \times \sqrt{N_{\rm pixel}},
\label{eq2}
\end{equation}
where the median $\langle f \rangle$ and standard deviation $\sigma(f)$ of flux densities are estimated within a 200$\rm \AA$-wide region, avoiding the effect of sky lines and emission lines. $\rm \sigma_G$ represents the width of the Gaussian profile, $ l_C$ represents the emission line center, $l'$ is the spectral dispersion ($l'$=1$\rm \AA/$spaxel), and $N_{\rm pixel}$ equals to $5\times \sigma_{\rm G} /l'$. After checking each spectrum, we find that spaxels closer to the northern star-forming region have stronger emission lines of ionized gas, including $\rm H\alpha$, $\rm H\beta$, $\rm [NII]\lambda \lambda6548,6583$ and $\rm [OIII]\lambda \lambda4959, 5007$, while the fluxes of $\rm H\beta$, $\rm [NII]\lambda \lambda6548,6583$ and $\rm [OIII]\lambda \lambda4959, 5007$ of the south region decrease significantly, which can be clearly seen in the spectra shown in Figure~\ref{fig2}. Below we only impose the emission lines with S/N $\ge$ 5. 

 \begin{figure*}
  \begin{center}
  \includegraphics[width=0.95\textwidth, angle=0 ]{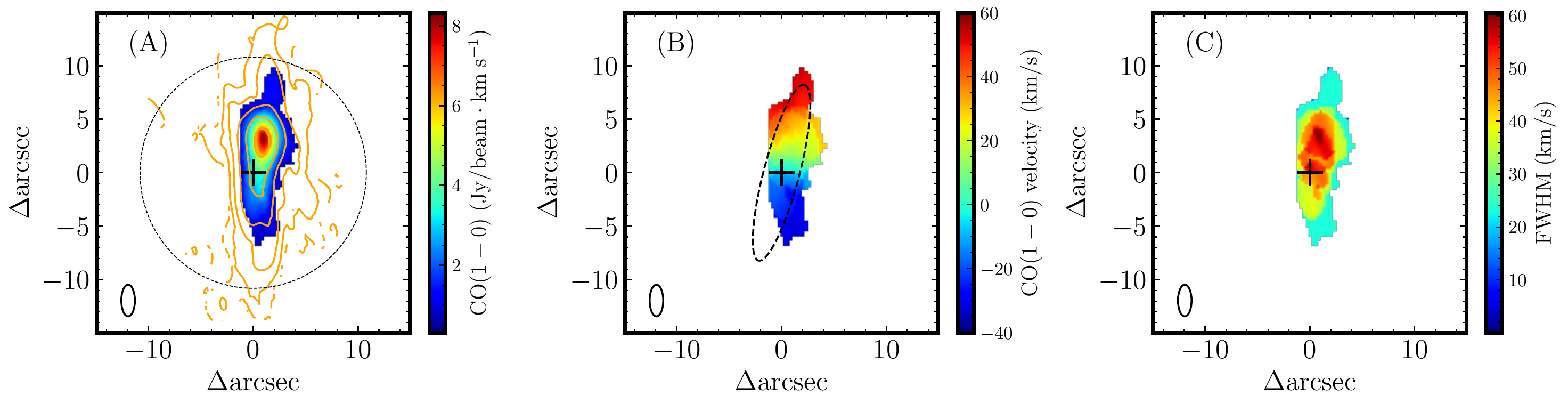}
  \end{center}
  \caption{Moment maps of the CO(1-0) emission from NOEMA using a threshold of 17.4 mJy ($\sim 5\sigma_{\rm rms}$). Panel A is the integrated intensity in Jy/beam $\cdot$ km/s, Panels B and C show the mean velocity and line width (FWHM) of CO(1-0), respectively. The orange contours show the distribution of the moment0 map with a threshold of 3.75 mJy,  covering a larger area ($\sim$2.4 kpc$^2$). The contours are corresponding to the values of 0.1, 0.2, 0.4, 0.8, 1.6, 3.2, and 6.4 Jy/beam$\rm \cdot$ km s$^{-1}$. The dashed black circle shows the primary beam size (21.6$\arcsec$) of IRAM 30-m telescope. The dashed black ellipse represents the position and length of stellar bar, as shown in Panel C of Figure~\ref{fig3}. The symbols of the black plus and left-bottom ellipse indicate the center of galaxy and the beam size ($\rm 2.\arcsec96 \times 1.\arcsec30$), respectively.} 
  \label{fig4}
\end{figure*}

Panels G and H of Figure~\ref{fig3} show the distributions of  the $\rm H\alpha$ velocity and $\rm H\alpha$ velocity dispersion, respectively. The rotation of ionized gas is consistent with that of stellar component. The average velocity dispersion of ionized gas in the central region is $\sim$140 km$\rm s^{-1}$. Considering the nebular extinction, $\rm H\alpha$ flux is corrected by assuming the ``Case B'' recombination model, (i.e, $\rm H\alpha /H\beta =2.86$, $\rm T=10^4 K$ and $n_{\rm e}=10^2 \rm cm^{-3}$) and the \cite{Calzetti+00} extinction law, which means that the corrected $\rm H\alpha$ flux should satisfy S/N$_{\rm H\alpha}>$5 and S/N$_{\rm H\beta}>$5 simultaneously. The distribution of the corrected  $\rm H\alpha$ flux is shown in Panel F of Figure~\ref{fig3}. For a better view, we present a zoom-in image in Panel I. Due to the weak $\rm H\beta$ emission near the southern star-forming region, few spaxels (S/N$_{\rm H\alpha, H\beta}>$5) in south are left and most corrected $\rm H\alpha$ emission concentrates on the northern blue star-forming region. Based on the extinction-corrected $\rm H\alpha$ emission flux, the extinction-corrected SFR can be derived by using the formula given by \cite{Kennicutt+1998a}:
\begin{equation}
\label{eq3}
{\rm SFR} \; [ {\rm M_\odot yr^{-1}}]=7.9\times10^{-42} L(\rm H\alpha),
\end{equation}
where $L(\rm H\alpha)$ is the extinction-corrected $\rm H\alpha$ luminosity.
Within the $\rm H\alpha$ region of Panel I, the estimated SFR is about 0.27$\pm$0.05 $\rm M_\odot yr^{-1}$.
After summing up spectra of all available 1792 spaxels in PGC 34107, the total SFR listed in Table~\ref{tab1} is estimated about 0.33$\pm$0.01 $\rm M_\odot yr^{-1}$, which is between the results from the MPA-JHU DR7 catalog and \cite{Zhou+18}. Their SFRs shown in Table~\ref{tab1} have been converted to the \cite{Salpeter+1955} IMF. Most of SFR is mainly contributed by the northern star-forming region. We adopt the axial ratio of the inner disk ($q=0.84$) from \cite{Zhou+20} to derive the inclination angle by
\begin{equation}
\label{eq4}
\cos^{2} i=\frac{q^2-q^{2}_{0}}{1-q^2_0},
\end{equation}
where $q_0=0.25$ for classical S0 galaxies (\citealt{Sandage+1970}). The derived inclination is estimated to be $i$=34$^\circ$. Thus, within the $\rm H\alpha$ region of Panel I, the inclination-corrected surface density of SFR is about 0.31$\pm$0.06 $\rm M_\odot yr^{-1} kpc^2$, which is listed in Table~\ref{tab1}. Overall, the IFU 2D-spectroscopic data clearly reveals that the star formation occurs near the central region of the galaxy, and is mainly concentrated in the northern region, which is $\sim$380 pc away from the galactic center.

 \begin{figure*}[!t]
  \begin{center}
  \includegraphics[width=0.95\textwidth, angle=0 ]{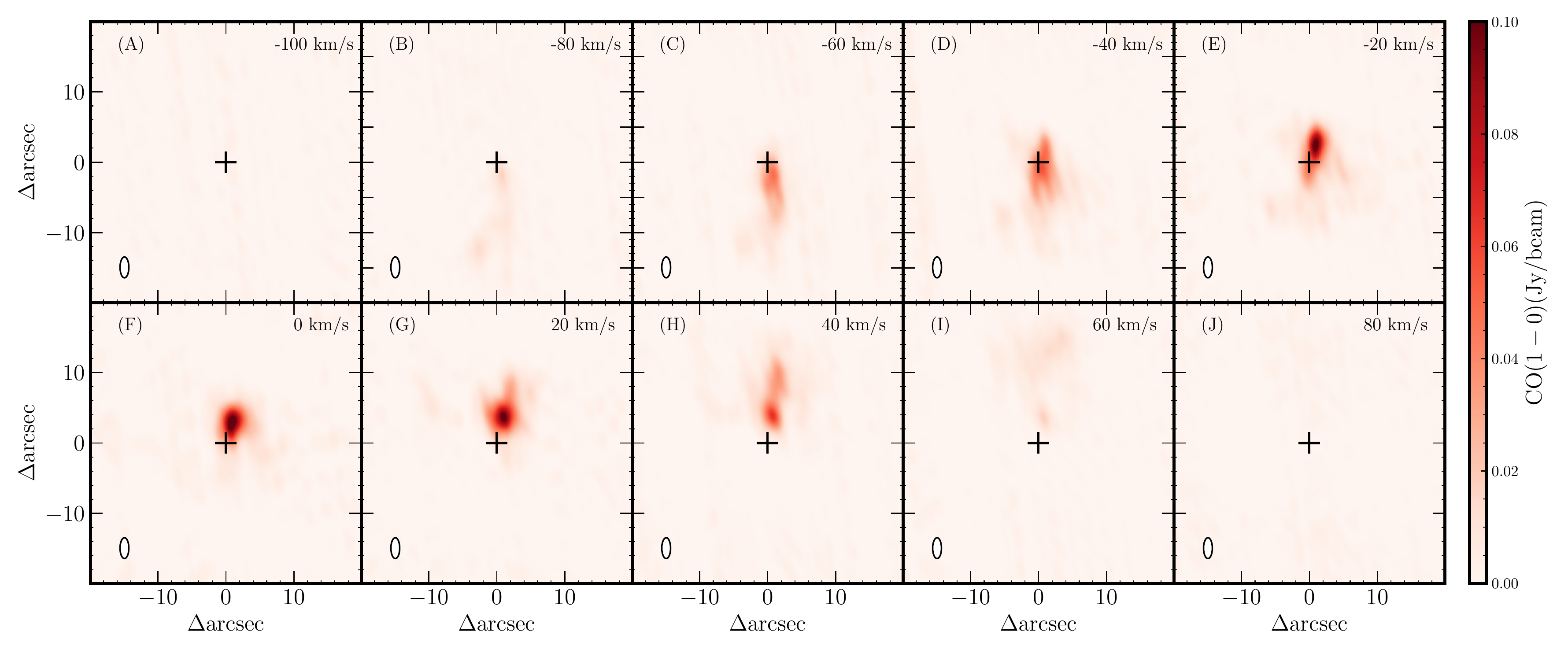}
  \end{center}
  \caption{Channel maps of the CO(1-0) emission with the velocity range -100 km/s $<v<$ 80 km/s and velocity interval $\Delta v=$20 km/s. The corresponding value of velocity is shown in the upper right corner. The beam size is marked with black ellipse in the bottom left corner. The intensity in all maps is scaled by the same level, as shown in color bar.}
  \label{fig5}
\end{figure*}
       
\section{NOEMA Millimeter Data}
\label{sect3:NOEMA}

\subsection{NOEMA Millimeter Observation}
\label{sect3.1:NOEMA Observation}

PGC 34107 was observed twice in the $^{12}$CO(1-0) transition, hereafter CO(1-0), with NOEMA on 2019 December 15th and 2020 January 3rd, respectively (project W19BJ; PI: Xue Ge). The observations were carried out  with ten antennas in the C configuration with 6.0-hr total on-source time. In the first 3-hr observation on 2019 December 15, 1055+018 was used as the receiver bandpass (RF) and phase/amplitude calibrators simultaneously, while 1038+064 was used as the absolute flux calibrator. In the second 3-hr observation on 2020 January 3rd, 1055+018 was used as the phase/amplitude calibrators, while the RF and the absolute flux calibrators were 3C273 instead. The antennas are equipped with dual polarizations for the 3 mm atmospheric window (93.4-116.0 GHz).  Each polarization covers a bandwidth of $\sim$7.8 GHz at a spectral resolution of 2 MHz. 

The calibrations of CO(1-0) were performed by using CLIC, including the RF, flux, phase and amplitude calibrations. Cleaning and imaging of data were done by using MAPPING. Both CLIC and MAPPING are modules of the available GILDAS\footnote{http://www.iram.fr/IRAMFR/GILDAS} software package. The redshifted CO(1-0) frequency is 114.731 GHz, given its redshift of $z$=0.00471 ($\rm \nu_{rest}=115.271$ GHz). After subtracting the continuum in the \texttt{uv} plane,  we image a CO(1-0) datacube, which covers a velocity range of -480$\sim$500 km/s at a velocity resolution of 20 km/s. The synthesized beam size is $\rm 2.\arcsec96 \times 1.\arcsec30$ with a position angle of 13.70$^{\circ}$. The final datacube consists of $\rm 512 \times 512$ pixels for a map cell of $\rm 0.\arcsec29 \times 0.\arcsec29$, covering a FoV of $\rm 43.\arcsec9 \times 43.\arcsec9$.  
In the ATLAS$^{\rm 3D}$ project, the Combined Array for Research in Millimeter Astronomy (CARMA) CO imaging survey of ETGs (\citealt{Alatalo+13}) also included PGC 34107, which was observed in 3.75 hours with CO-synthesized descriptionbeam of $3.\arcsec8 \times 3.\arcsec3$ and pixel of $1\arcsec \times 1\arcsec$. Compared to the CARMA CO observation, 
our NOEMA observation has a longer exposure time and smaller beam size, which is helpful to find the gas inflow along the stellar bar (see Section~\ref{sect4.1:gas inflow}).

Besides CO(1-0), we also find a strong isotopic line of $^{13}$CO(1-0) in the upper inner baseband, which was detected by \cite{Crocker+12}, using IRAM 30-m telescope at Pico Veleta, Spain. We use CLIC and MAPPING to clean and image the $^{13}$CO(1-0) datacube. The redshifted (rest) $^{13}$CO(1-0) frequency is 109.685 (110.201) GHz. The synthesized beam size is $\rm 3.\arcsec07 \times 1.\arcsec36$ with a position angle of 13.47$^{\circ}$. The final $^{13}$CO(1-0) datacube is made up of $\rm 512 \times 512$ pixels for a map cell of $\rm 0.\arcsec31 \times 0.\arcsec31$, covering a FoV of $\rm 45.\arcsec9 \times 45.\arcsec9$. The integrated NOEMA fluxes are in the unit of Jy $\rm km s^{-1}$, including the CO(1-0) and  $^{13}$CO(1-0).

\subsection{CO(1-0) Distribution}
\label{sect3.2:CO(1-0)}     

We use a threshold of 17.40 mJy ($\sim$5$\sigma_{\rm rms}$) to output moment maps of the CO(1-0) emission. Figure \ref{fig4} shows the moment maps of CO(1-0) emission, including the integrated intensity (Panel A), velocity (Panel B) and velocity dispersion (Panel C), respectively. Most of the CO emission concentrates on the central region, but its flux peak is off-center and closer to the northern star-forming region, similar to $\rm H\alpha$ emission.
We measure the PAs of CO(1-0) and stellar velocity fields by using IDL package KINEMETRY (\citealt{Krajnovic+06}), where the kinematic PA is defined as the counterclockwise angle between north and a line which the velocity field of gas or stars on the redshifted side. The PA of molecular gas CO(1-0) and stars with $3\sigma$ error is $\rm PA_{mol}$=12.9$^{\circ}\pm$38.6$^{\circ}$ and $\rm PA_{star}$=12.9$^{\circ}\pm$6.4$^{\circ}$, respectively. There is no kinematic misalignment between molecular gas and stars, which is consistent with \cite{Davis+11}. The misalignment between molecular gas and stellar bar ($\rm PA_{bar}$=-15.97$^{\circ}\pm$0.03$^{\circ}$) is about $\rm \Delta$PA(mol, bar)=28.87$^{\circ}\pm$38.6$^{\circ}$. The molecular gas of PGC 34107 is mainly concentrated in a small central region, which makes the derived PA have a large error. Considering the error, the molecular gas could be aligned with the stellar bar, as shown in Panel B of Figure~\ref{fig4}. In Panel C, the velocity dispersion shows an increase from the outside to inside, which may be caused by the gas turbulence near the star-forming region. Here, we have adopted the $\rm ^{3D}$BAROLO code (\citealt{Di+Teodoro+15}) to perform a 3D-fitting on the CO(1-0) data cube to check the effect of beam smearing on the velocity dispersion. Figure~\ref{fig5} shows the channel maps of CO (1-0) with a velocity range from -100 to 80 km/s and a velocity interval of 20 km/s. It can clearly find that CO(1-0) emissions are concentrated on the center and its peak offsets to the north, $\rm \sim 380$ pc away from the center. Along the major axis of galaxy, a regular rotation of CO (1-0) is dominated.

\begin{figure}[!t]
  \centering
  \includegraphics[width=0.9\textwidth, angle=0 ]{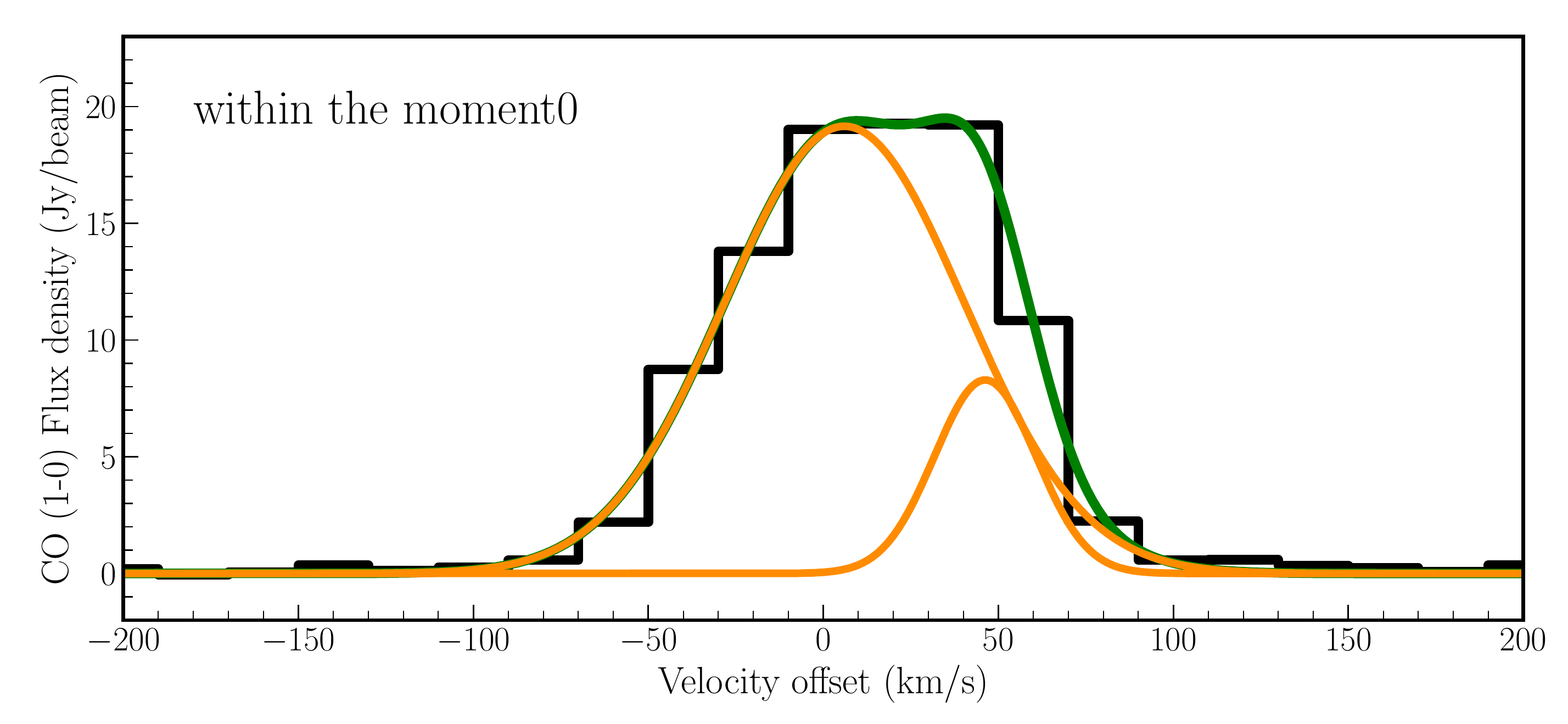}
  \caption{Decomposition of the CO(1-0) line profile with a velocity interval of 20 km/s. The black line represents the original stacked spectrum within the moment0 map, and the orange lines represent Gaussian models. The green line is the superposition of the two Gaussian profiles as the best-fit model.}
  \label{fig6}
\end{figure}

\begin{figure*}[!t]
  \begin{center}
  \includegraphics[width=0.8\textwidth, angle=0 ]{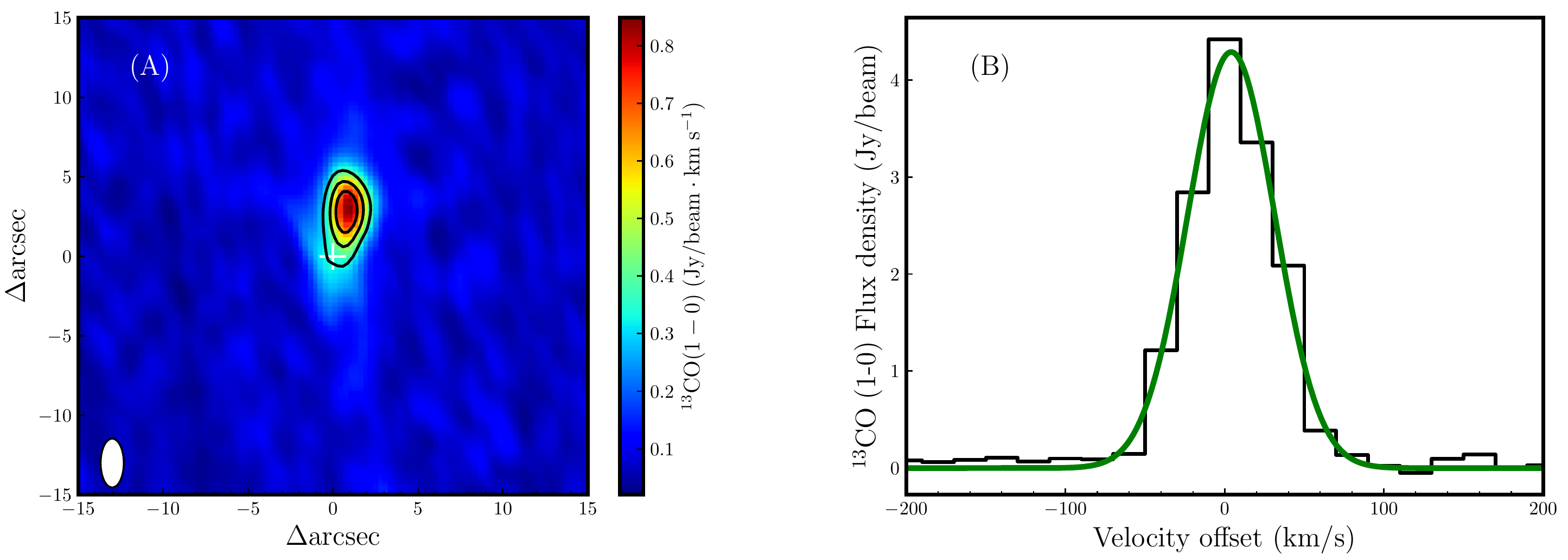}
  \end{center}
  \caption{Panel A: The $^{13}$CO(1-0) integrated intensity map, overlaid with black contours (corresponding to $\sim$40\%, 60\% and 80\% of the peak). The optical center is marked with a white cross. The white ellipse indicates the corresponding beam size ($\rm 3.\arcsec07 \times 1.\arcsec36$) of $^{13}$CO(1-0). Panel B: The stacked $^{13}$CO(1-0) line profile with a velocity interval of 20 km/s by using a threshold of 6.5 mJy ($\rm > 5\sigma_{rms}$). The green line represents the best-fit Gaussian profile.}
  \label{fig7}
\end{figure*}

We extract the CO emission of each pixel within the moment0 map, i.e., Panel A of Figure~\ref{fig4},  to construct a stacked spectrum, as shown in black histogram of Figure~\ref{fig6}. We find that the CO(1-0) line shows asymmetric profile, and fit the spectrum with two Gaussian components shown in orange solid lines. One stronger component is the primary component of the rotation of molecular gas, while the other is the residual due to the asymmetry of the distribution and the difference of flux intensity for extracted pixels.This secondary component might give a hint that the molecular gas may have weak signatures of non-circular motion, which will be discussed in Section~\ref{sect4.1:gas inflow}. The best-fit line is the superposition of them shown in blue solid line. Within the CO(1-0) coverage of Figure~\ref{fig4}, we also estimated the luminosity of molecular gas by using the relation from \cite{Solomon+05}: 
\begin{equation}
\label{eq5}
   L'_{\rm CO} [{\rm K\; km\; s^{-1}}]= 3.25\times 10^{7} S_{\rm CO}\Delta v \nu_{obs}^{-2} D_{L}^{2} (1+z)^{-3},
\end{equation}
where $S_{\rm CO}\Delta v$ is the CO integrated flux density in unit of $\rm Jy \,km \, s^{-1}$, $\nu_{obs}$ is the observed frequency in GHz at a given redshift, and the {\bf luminosity distance} is $D_{\rm L}$=20.2 Mpc. The CO luminosity is about $\rm (3.80\pm0.26)\times10^{7}\; \rm K \;km \; s^{-1} pc^2$, shown in Table~\ref{tab1}. The corresponding mass of the molecular hydrogen ($\rm M_{ H_2}$) can be estimated by adopting the conversion factor ($\rm \alpha_{CO}$) between CO(1-0) and $\rm H_2$:
\begin{equation}
\label{eq6}
    {\rm M_{H_2}} = \alpha_{\rm CO} \times L'_{\rm CO},
\end{equation}
where $\alpha_{\rm CO}$ is adopted to be 4.3 $\rm (M_\odot km\; s^{-1} \;pc^2)^{-1}$ for the inner disk of our galaxy (\citealt{Bolatto+13}). The molecular hydrogen mass ($\rm M_{ H_2}$) is estimated to be $\rm (1.66\pm0.50)\times10^{8} M_\odot$ within the moment0 map ($\sim$0.63 kpc$^2$) at $D_{\rm L}$=20.2 Mpc. Using the IRAM 30-m telescope, \cite{Young+11} obtained the $\rm M_{ H_2}$ of $\rm 1.90\times 10^8 M_\odot$ at $D_{\rm L}$=20.2 Mpc. 
It is consistent with the result from \cite{Young+11}, considering the 10\% absolute flux uncertainties for NOEMA at 3mm. Covering a larger moment0 map ($\sim$2.51 kpc$^2$) where some pixels extend beyond the primary beam of IRAM 30-m telescope, \cite{Alatalo+13} derived the $\rm M_{ H_2}$ of $\rm 2.29\times10^{8} M_\odot$ at $D_{\rm L}$=20.2 Mpc. 
Under a similar coverage (2.4 $\rm kpc^2$, orange contours in Panel A of Figure~\ref{fig4}), we estimate that the $\rm M_{ H_2}$ is $\rm (2.51\pm0.77) \times 10^{8} M_\odot$, which agrees with \cite{Alatalo+13}. Note that here $\rm M_{ H_2}$ from different work has been converted by adopting the same $\alpha_{\rm CO}$ and $D_{\rm L}$ as this work. In summary, our estimated $\rm M_{ H_2}$ is consistent with previous work. Below we adopt the estimated $\rm M_{ H_2}$ within $\sim$0.63 kpc$^2$ at $D_{\rm L}$=20.2 Mpc to discuss. The inclination-corrected surface mass density of $\rm H_2$ is about $\rm (2.19\pm 0.65)\times 10^{2} M_\odot \; pc^{-2}$. We calculate the star formation efficiency (SFE) by the ratio of SFR within the $\rm H\alpha$ region in Panel I of Figure~\ref{fig3} to the $\rm M_{H_2}$ of $\rm (1.66\pm0.50)\times10^{8} M_\odot$. The SFE is about $\rm (1.63\pm 0.57)\times 10^{-9} yr^{-1}$, which is similar with that of spirals ($\rm 1.5\times 10^{-9} yr^{-1}$, \citealt{Kennicutt+1998b}) but higher than that of ETGs ($\rm 4\times 10^{-10} yr^{-1}$, \citealt{Davis+15}). Assuming a constant consumption of gas, the depletion time is about 600 Myr.

\subsection{$^{13}$CO(1-0) Distribution}
\label{sect3.3:13CO(1-0)}

We integrate the flux of all pixels with signals to obtain the intensity diagram of $^{13}$CO(1-0) isotopes, as shown in Panel A of Figure~\ref{fig7}. The resolved $^{13}$CO(1-0) map manifests that most of $^{13}$CO isotopes throughout the galaxy mainly distribute in the northern active star-forming region. We sum all spectra of $^{13}$CO(1-0) pixels by using a threshold of 6.5 mJy ($\rm > 5\sigma_{rms}$) to construct a stacked spectrum of $^{13}$CO(1-0), shown as black histogram in Panel B of Figure~\ref{fig7}. There is no significant asymmetry for $^{13}$CO(1-0) profile, so we fit it with a single Gaussian. The best-fit line is shown as green. The estimated $^{13}$CO(1-0) integrated intensity ($\rm I_{13CO}$) and luminosity ($L'_{\rm 13CO}$) are $5.35\pm 0.20$ Jy km$\rm s^{-1}$ and ($0.58\pm 0.02$) $\times 10^7$ K km $\rm s^{-1} pc^{2}$, respectively.

\section{Discussion}
\label{sect4:Discussion}
\subsection{The Inflow of Molecular Gas CO(1-0)?}
\label{sect4.1:gas inflow}

\begin{figure*}[!t]
  \begin{center}
  \includegraphics[width=0.7\textwidth, angle=0 ]{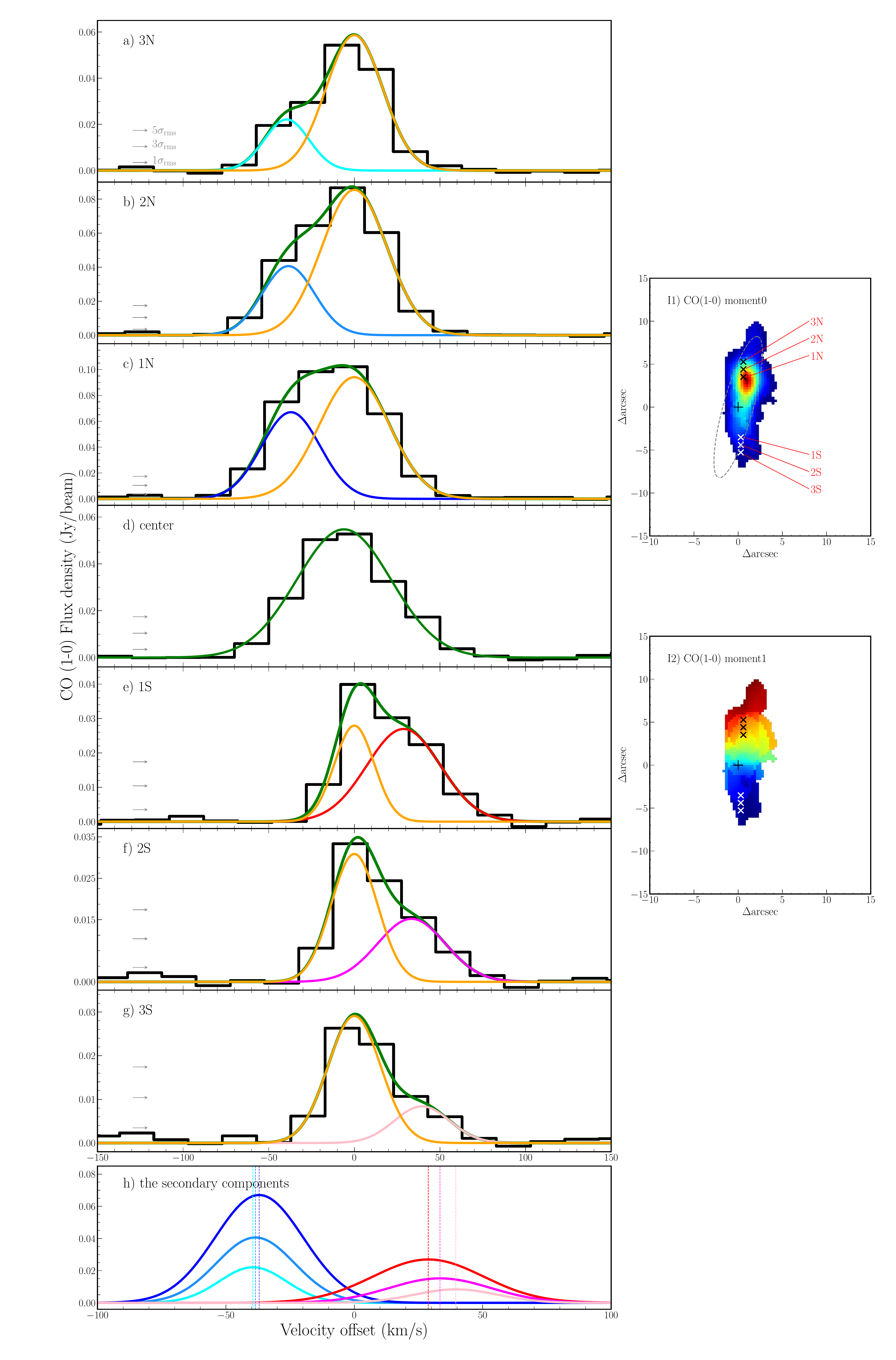}
  \end{center}
  \caption{\small CO(1-0) spectra for seven extracted pixels along the bar. The location of the seven pixels are marked and shown in CO(1-0) flux and velocity map of right Panels I1) and I2). The pixels on the north and south sides are selected along the stellar bar with equal spacing. The bar's position and length is shown in dotted gray ellipse in right Panel I1). The black histograms in Panel a) to g) are the original CO(1-0) spectra from the 3N, 2N, 1N, central, 1S, 2S, and 3S pixels as shown in Panel I1). Note that the local velocity of each original spectrum is subtracted.
 Except the center one, each spectrum is fitted by utilizing two gaussian profiles. One is the primary gaussian profile in orange, the other is the secondary gaussian profile in blue for the blueshifted component or in red for the redshifted component, respectively. As the distance approaches the center, the colors of blue and red lines get darker. The green lines are the superposition of the two gaussian profiles as the best-fitting. The bottom panel h) shows all  those secondary profiles, whose centers are marked with dashed vertical lines in corresponding colors. The gray arrows in Panel a) to g) represent the positions of $\rm \sim 5\sigma_{rms}$, $\rm \sim 3\sigma_{rms}$ and $\rm \sim 1\sigma_{rms}$ are, respectively. 
 The S/Ns of all secondary components are $\rm > 3\sigma_{rms}$, expect the 3S spectrum ($\rm > 2\sigma_{rms}$).}
  \label{fig8}
\end{figure*}

\begin{figure*}
  \begin{center}
  \includegraphics[width=0.65\textwidth, angle=0 ]{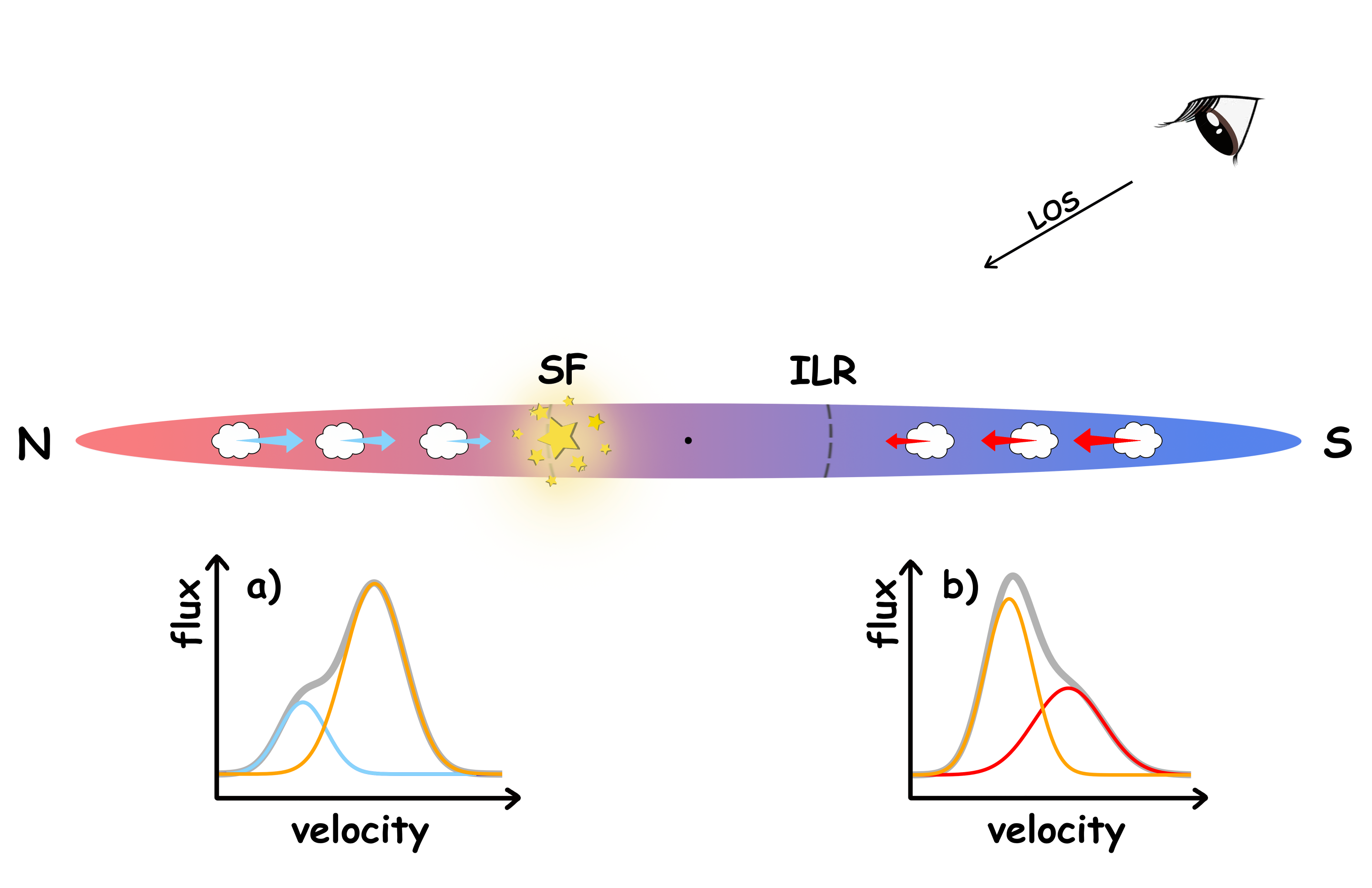}
  \end{center}
  \caption{The sketch of molecular gas inflow along the stellar bar of PGC 34107. The stellar bar is represented by an elliptical shape, with its receding and approaching sides indicated in red and blue, respectively. White clouds, golden stars and dashed lines represent the molecular gas, star formation (SF) region and inner Lindblad resonances (ILR), respectively. The basic directions are also shown as N and S in the plot. The gas-infalling velocity (receding/approaching side: blue/red arrows) decreases with the decreasing distance to the galaxy center, as shown by shrinking arrows. The line profile in Panel a (b)  is corresponding to the 1-3N (1-3S) line profiles in Panels a-c (e-g) of Figure~\ref{fig8}. In a the line-of-sight (LOS) direction, a blueshifted component (Panel a: blue line) and a redshifted component (Panel b: red line) can be observed, compared to the local circular rotations (orange lines).} 
  \label{fig9}
\end{figure*}

The high quality of NOEMA gives us a chance to identify weak signatures of non-circular motion. 
As shown in right Panel  I1) of Figure~\ref{fig8}, we separately extract 3-pixel molecular spectra at equal spacing along the bar on the northern and southern sides of the galaxy, which are marked with 3N, 2N, 1N, and 1S, 2S, 3S, respectively. Here we consider that the space of pixels is symmetric along the stellar bar, and each spectrum possesses an enough high S/N (in gray arrow) to obtain a reliable secondary component, simultaneously. From Panels a) to g) of Figure~\ref{fig8}, we show the corresponding observed CO(1-0) line profiles in the north, center and south of galaxy with black histograms, respectively. The asymmetry of the line profiles can be found, thus we adopt two Gaussian profiles to fit the observed CO profiles except the central one. The orange lines represent the primary components for the local rotation of molecular gas, while the lines in a series of blue (red) colors represent the secondary velocity components, which are corresponding to the blueshifted (redshifted) parts of the local CO line profiles. Here the centroid of the primary Gaussian component of each observed CO(1-0) spectrum has been subtracted in order to compare the blueshifted and redshifted secondary components clearly. The S/Ns of all secondary components are $\rm > 3\sigma_{rms}$, expect the 3S spectrum ($\rm > 2\sigma_{rms}$) in Panel g). The darker the color, the closer it is to the center of galaxy. The green line is the superposition of the two Gaussian profiles. In the bottom panel h) of Figure~\ref{fig8}, all secondary components in blue and red are shown together. 

In Figure~\ref{fig8}, we find that three pixels in the north (i.e., 3N, 2N and 1N) are corresponding to the receding side of the galaxy seen in Panel I2) of Figure~\ref{fig8}, while their secondary gaussian components are blueshifted. It is inverse in three southern pixels. Three southern pixels in the approaching side of the galaxy exhibit redshifted secondary components, compared to their local CO line profiles, suggesting that there might be an inflow of the molecular gas along the bar. The location of those extracted pixels is within the radius (8$\arcsec$-10$\arcsec$, $<$ 1kpc) of the bar (\citealt{Zhou+20}). Panel h) of Figure~\ref{fig8} clearly shows that as the pixel approaches the center of the galaxy, the velocity of gas inflow decreases and the flux intensity increases. Previous studies of barred galaxies have found for decades that the linear barlike gas morphology in the central regions does not last all the way into the nucleus (\citealt{Ishizuki+1990,Kenney+1992}). It indicates the inflow may be slowed down or stopped at certain radii, such as inner Lindblad resonances (ILR, \citealt{Combes+1988, Shlosman+1989, Combes+01}), which are located inside the bar with a radius of a few hundred parsec as seen in CO(1-0) (\citealt{Kenney+1992, Schinnerer+00,Walter+02, Olsson+10}). The location of our extracted pixels within $\sim$300-550 pc is connected to the peak of the star formation ($\sim$ 380 pc), and the position of the flux peak might be related to ILR, which seemingly provides a reliable explanation for why the velocity of gas inflow decreases gradually with the decreasing distance, finally resulting in an active star formation. 

Based on the observations of the CO(1-0) line made with the Berkeley-Illinois-Maryland Association (BIMA) millimeter array, \cite{Regan+1999} studied the kinematics of the dense molecular gas in a set of seven barred spiral galaxies. By extracting the spectrum along the bar, they found the two velocity components near the joint of dust lanes and nuclear ring. The one velocity component is corresponding to gas on circular orbits, the other one is attributed to the gas flowing down the bar dust lane, which is similar to what we found here. This implies that it is easier to detect gas inflows at such joint locations. For PGC 34107, we have tried to adopt $\rm ^{3D}$BAROLO software to perform a 3D tilted-ring fitting on the CO(1-0) emission line data cube (\citealt{Di+Teodoro+15}), considering the variation of inclination, PA, rotation velocity and velocity dispersion (see the Appendix for detail). The position-velocity diagrams along major and minor axis indicate the existent of a weak non-circular motion compared to circular model. Generally, the contribution of the radial inflow motions is small (\citealt{Di+Teodoro+21}),
while our existing velocity and spatial resolutions are not high enough to resolve the weak no-circular motion completely, and a higher resolved observation is needed in the future. Thus, here we explore the possibility that the observed velocity structure is due to radial inflow.

Assuming that the non-circular motion is due to the bar-induced gas inflow, we plot a simple schematic view about the kinematics and spectra in the stellar bar (elliptical shape) of PGC 34107, shown in Figure~\ref{fig9}. The filled red/blue colors imply the receding/approaching rotation in the north/south. As the gas (white cloud) is closer to the inner Lindblad resonances (ILR, dashed lines), its velocity (blue/red arrows) gradually decreases. The gas gradually slows down and accumulates near the ILR, providing raw materials for star formation (SF). Panels a/b of Figure~\ref{fig9} show the spectral sketch of molecular gas along a line-of-sight (LOS) direction. The spectrum in Panel a) is corresponding to the 3N, 2N and 1N spectra in Panel a)-c) of Figure~\ref{fig8}, while Panel b) is for the 1S, 2S and 3S ones in Panels e)-g) of Figure~\ref{fig8}. Compared to the local circular rotation (orange lines), the blueshifted/redshifted components (blue/red lines) can be observed on the receding/approaching sides of stellar bar in PGC 34107. In Figure~\ref{fig8}, the flux intensity for secondary component is higher in the north than that in the south, which suggests that the gas inflow in the north is stronger than that in the south. It might explain why the star formation is concentrated at the northern side. (see the SF region of Figure~\ref{fig9}).
The hydrodynamics simulations of barred galaxies suggested that thin bar with the axis ratio (b/a) of $\sim$ 0.2 can produce the mass inflows of 0.25 $\rm M_\odot$yr$^{-1}$ into the inner $\sim$100pc (\citealt{Piner+1995}). PGC 34107 has a thin bar with the axis ratio of 0.24 derived by \cite{Zhou+20}. We convert the Equation 4 of the mass outflow rate from \cite{Gracia-Burillo+14} to simply derive the mass inflow rate as: 
\begin{equation}
\label{eq7}
\frac{dM}{dt}=3\times V_{\rm in}\times \frac{M_{\rm mol}}{R_{\rm in}} \times \tan \alpha,
 \end{equation}
 where $V_{\rm in}$ and $R_{\rm in}$ are the projected velocity and radial size of inflow, respectively. The $M_{\rm mol}$ represents the molecular gas of inflow. From Panel h) of Figure~\ref{fig8}, the mean gas-infalling velocity is about 35 km $\rm s^{-1}$. The $M_{\rm mol}$ is derived by summing the secondary components at these six pixels in north and south.
Assuming an angle between inflow and the line-of-sight of $40^{\circ}$, and a projected radial size of 550 pc (corresponding to the position of 3N or 3S pixel), the mass inflow rate ($dM/dt$) is about 0.11 $\rm M_\odot yr^{-1}$, which is in the same order of magnitude as \cite{Piner+1995}. Note that although in a barred galaxy the non-circular motions are elliptical streaming motions induced by the bar, the actual inflow is a small fraction of the non-circular velocity and difficult to measure directly (\citealt{Di+Teodoro+21}). So the assumed $V_{\rm in}$ is overestimated and the current $dM/dt$ is the upper limit based on existing observations and assumptions.

\subsection{ Gas Distributions: $\rm H\alpha$, CO(1-0) and $^{13}$CO(1-0)}
\label{sect4.2:gas distributions}

\begin{figure}
   \centering
  \includegraphics[width=0.8\textwidth, angle=0 ]{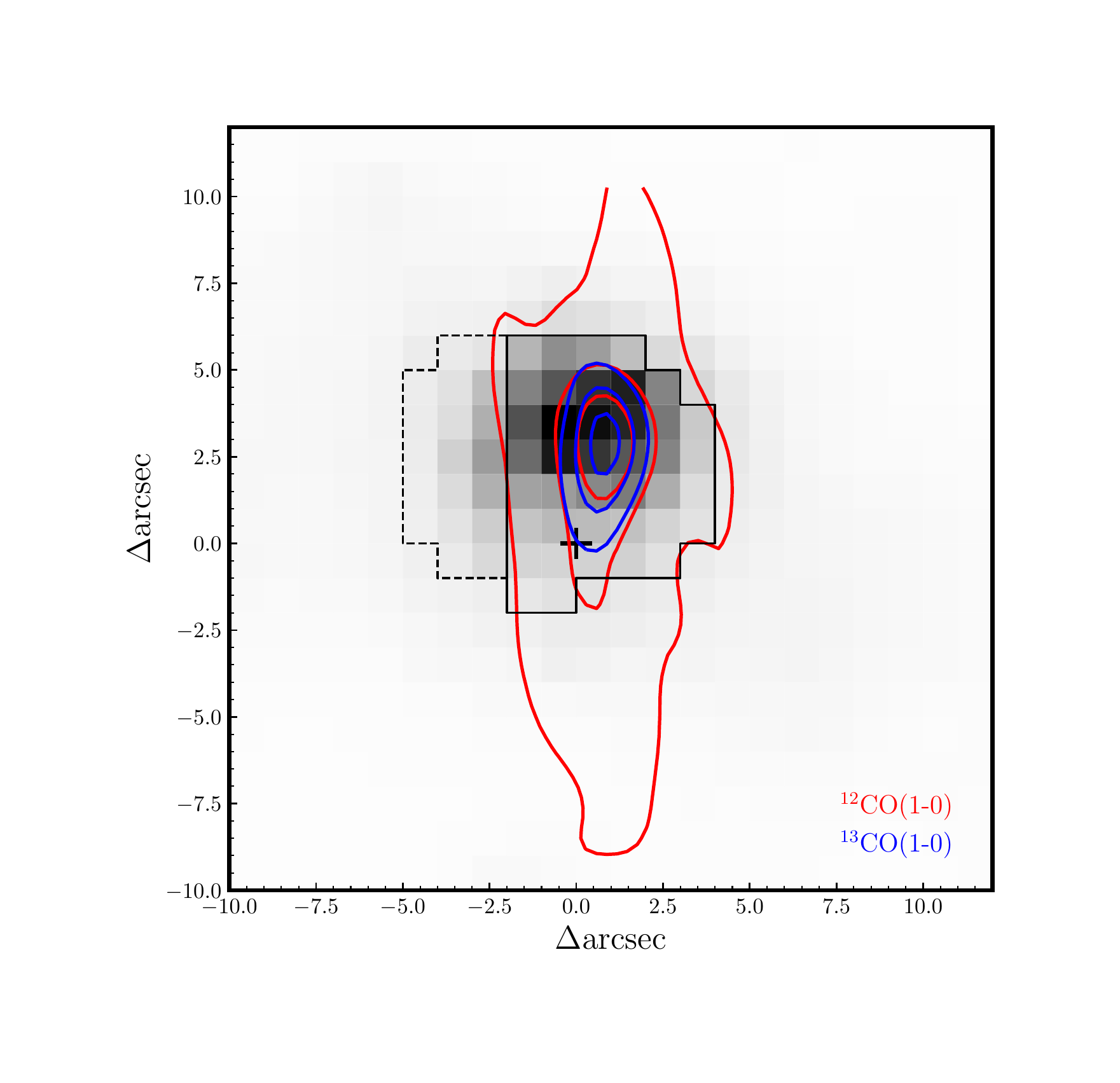}
  \caption{The illustration of the distributions of $\rm H\alpha$, CO(1-0) and $^{13}$CO(1-0) in PGC 34107. The greyscalerepresents the distribution of non-corrected $\rm H\alpha$ flux while corrected $\rm H\alpha$ region is shown in the (dashed +solid) black polygon, same as Panel I in Figure~\ref{fig3}. The blue contours represent the $^{13}$CO(1-0) integrated intensity, same as in Panel A of Figure~\ref{fig7}. The red contours is corresponding the 10\%, 40\% and 70\% of the CO(1-0) flux peak. The polygon covered by corrected $\rm H\alpha$ and CO(1-0) simultaneously is displayed in a solid black line ($\sim$40arcsec$^2$). The optical center is marked with a black cross.}
  \label{fig10}
\end{figure}

Since the spatially resolved optical and millimeter maps have been obtained by CAHA and NOEMA, respectively, we can overlay the CO(1-0) and $^{13}$CO(1-0) over a non-corrected $\rm H\alpha$ mosaic map to compare their different distributions, as shown in Figure~\ref{fig10}. The previous spatially resolved optical data didn't cover the longer wavelength (\citealt{Cappellari+11}), so in this work we could combine the resolved $\rm H\alpha$ with CO together.
It clearly shows that most of ionized and molecular gas distribute at the central region, especially concentrated on the northern star-forming region, while the $^{13}$CO isotopes mainly assemble in the inner area of star formation. 
\cite{Davis+TA+14} firstly reported that there was a strong positive correlation between the $^{12}$CO(1-0)/$^{13}$CO(1-0) intensity ratio (hereafter $\Re_{10}$) and star formation rate surface density ($\rm \Sigma_{SFR}$) in nearby galaxies. The corresponding Spearman's rank correlation coefficient was 0.755. They discussed that the causes of $\Re_{10}$ changes may be due to the systematically higher mean gas temperature and/or velocity dispersion in higher $\rm \Sigma_{SFR}$.
We estimate that the line ratio $\Re_{10}$ for the whole galaxy PGC 34107 is 7.52$\pm$0.01, which is consistent with that (7.5$\pm$1.2) of PGC 34107 observed by IRAM 30-m telescope from \cite{Crocker+12}. This line ratio is generally consistent with the ranges observed in spirals (\citealt{Paglione+01,Crocker+12}), but lower than the average value of 17.7$\pm$2.7 for starburst galaxies (\citealt{Mendez-Hernandez+20}). If we only use the pixels within the $\rm ^{13}$CO(1-0) blue contours in Figure~\ref{fig9} where the SFR is higher, the $\Re_{10}$ will become higher up to 11.2$\pm$0.7. Both corresponding velocity dispersion of CO(1-0) shown in Panel C of Figure~\ref{fig4} and the $\rm H\alpha$ flux are higher as well, which is seemingly consistent with the result of \cite{Davis+TA+14}. Generally, the most ionized and molecular gas concentre on the northern star-forming region, and the line ratio $\rm \Re_{10}$ is related to the degree of the star formation activity.

\subsection{MZR and SFMS}
\label{sect4.3: MZR and SFMS}

\begin{figure}
 \centering
 \includegraphics[width=0.8\textwidth, angle=0 ]{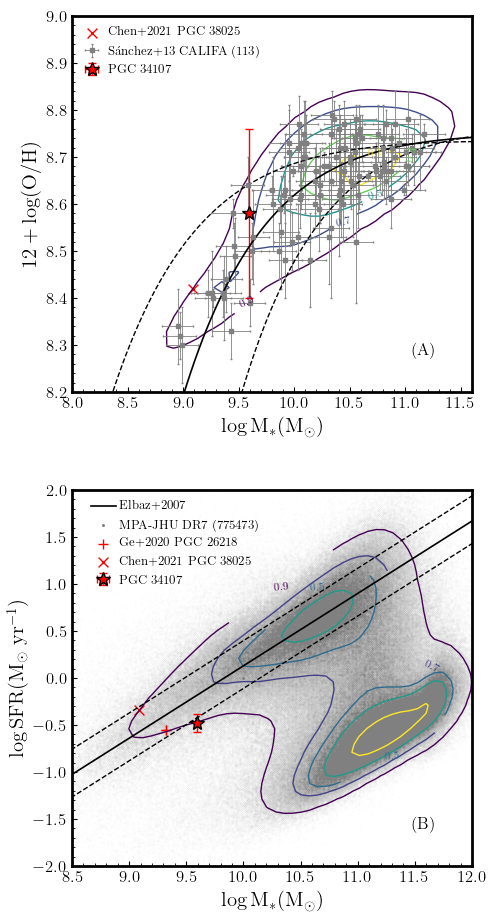}
 \caption{Panel A: The relation between gas-phase metallicity and stellar mass (MZR). The countours and grey dots represent the sample of 113 CALIFA galaxies from \cite{Sanchez+13}, where the gas-phase metallicity was also derived by O3N2 index. The black dashed and solid lines are from \cite{Sanchez+13} as well. 
Panel B: The star-forming main-sequence relation (SFMS). The contours and grey dots represent the sample of 775473 galaxies from MPA-JHU DR7 catalog (\citealt{Kauffmann+03, Brinchmann+04}). The SFMS for local SFGs is derived by \cite{Elbaz+07}, shown in black solid and dashed lines. Note that the stellar masses and star formation rates have been reconverted according to the \cite{Salpeter+1955} IMF instead the original \cite{Kroupa+01} IMF for the sake of consistency in this work. The symbols of red plus and multiplication represent the SFS0 PGC 26218 from \cite{Ge+20} and SFS0 PGC 38025 from \cite{Chen+21}, respectively. The red stars represent our target SFS0 PGC 34107. }
  \label{fig11}
\end{figure}

Panel A of Figure~\ref{fig11} shows the relationship between gas-phase metallicity and stellar mass, i.e., MZR, which is a sensitive diagnostic of galaxy evolution, including the gas inflow, metal production by stars, and outflows by galactic winds. Inflows can not only arrange the metallicity of a galaxy, but also provide the raw material for star formation. 
By stacking the spectra of all pixels in PGC 34107, we can also derive the global gas-phase metallicity.
Here the gas-phase metallicity is calculated by the linear relation between oxygen abundance and the O3N2 method, modified by \cite{Marino+13} as: 
\begin{equation}
\label{eq8}
    \rm 12+\log(O/H)=8.533-0.214\times O3N2,
 \end{equation}
 where $\rm O3N2=\log(\frac{[OIII]\lambda5007}{H\beta} \times \frac{H\alpha}{[NII]\lambda6583})$ was first introduced by  \cite{Alloin+1979}, after considering the extinction correction. Given the intrinsic scatter (0.18 dex) of this linear relation from \cite{Marino+13}, the global gas-phase metallicity and its scatter ($ 8.58 \pm 0.18$) are plotted in Figure~\ref{fig11}.
Based on IFS data provided by the CALIFA survey, \cite{Sanchez+13} explored the MZR of local spirals and adopted an asymptotic function to fit it. Their CALIFA galaxies are shown in contours and grey squares, while the best-fitting MZR is shown in black solid and dashed lines in Panel A of Figure~\ref{fig11}. Note that \cite{Sanchez+13} calculated the gas-phase metallicity based on the relation ($\rm 12 + \log(O/H) = 8.73 -0.32 \times O3N2$) from \cite{Pettini+Pagel+04}. This relation was updated to Function~(\ref{eq8}) by \cite{Marino+13}. Panel A of Figure~\ref{fig11} suggests that PGC 34107 follows the stellar mass-metallicity relation of local spirals. 
 
The global star-forming main sequence of PGC 34107 is shown in Panel B of Figure~\ref{fig11}. The background grey dots and contours represent the 775473 galaxies from MPA-JHU DR7 catalog (\citealt{Kauffmann+03, Brinchmann+04}), which provided the SFR and stellar mass based on a \cite{Kroupa+01} IMF, whereas we adopt a \cite{Salpeter+1955} IMF in this work. Therefore, we convert the SFR and stellar mass from MPA-JHU DR7 catalog by multiplying them by a factor of 1.5 (\citealt{Bell+03}). The black solid and dashed lines represent the local SFMS from \cite{Elbaz+07}, who obtained SFR and stellar mass from the MPA-JHU DR4 catalog. The SFS0 PGC 26218 (\citealt{Ge+20}) and PGC 38025 (\citealt{Chen+21}) are marked with the symbols of red plus and multiplication, respectively. Similar to the other two SFS0s, PGC 34107 basically follows the local SFMS relation within the error. Combining the MZR and SFMS with the plausible gas inflow for PGC 34107, it seemingly supports that the bar-induced gas inflow from the galactic disk supplies the raw material for the central star formation in PGC 34107. This could explain why the MZR and SFMS of PGC 34107, a lenticular galaxy, are comparable to spiral or star-forming galaxies.

\subsection{K-S law and MGMS}
\label{sect4.4: K-S law}

\begin{figure}
\centering
 \includegraphics[width=0.8\textwidth, angle=0 ]{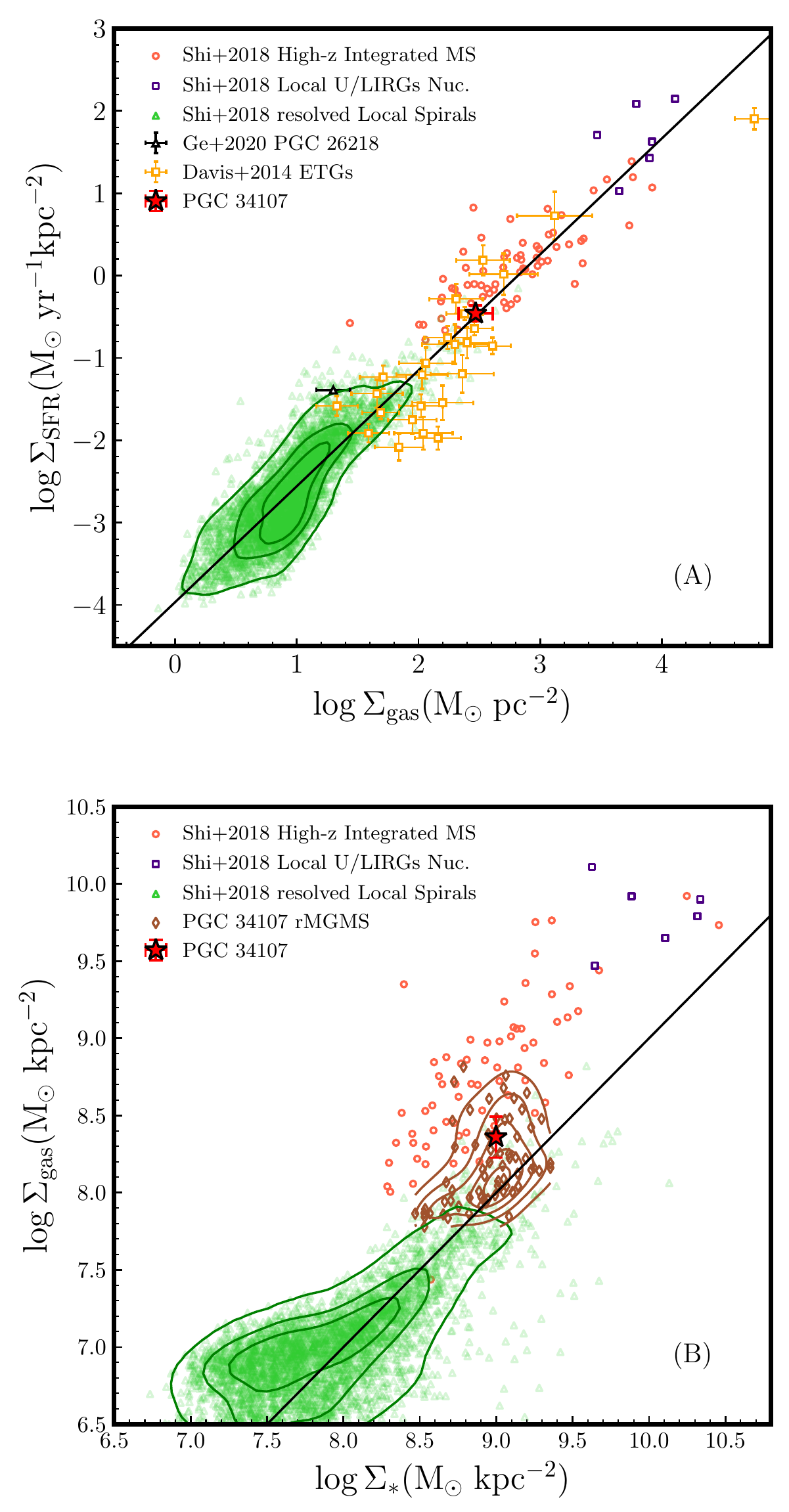}
 \caption{Panel A: The relationship between $\rm \log \Sigma_{gas}$  and $\rm \log \Sigma_{SFR}$, i.e., the Kennicutt-Schmidt (K-S) law. The red circles, indigo squares and green triangles/contours represent the high-z integrated main-sequence galaxies, the nuclear regions of local U/LIRGs and 12 resolved local spirals from \cite{Shi+18}. The black triangle and orange squares are the SFS0 PGC 26218 of \cite{Ge+20} and ETGs of \cite{Davis+14}. The solid line is the line of best fit for the \cite{Shi+18} sample. The red star represents our target SFS0 PGC 34107, calculated in a same coverage of CO(1-0) and $\rm H\alpha$ emission lines. Note that the $\Sigma_{\rm SFR}$ from different works has been converted to be based on the \cite{Salpeter+1955} IMF, and the $\Sigma_{\rm gas}$ has been corrected by adopting the same $\alpha_{\rm CO}$ in this work. Panel B: The relationship between $\rm \log \Sigma_{*}$ and $\rm \log \Sigma_{gas}$, i.e., the Molecular Gas Main Sequence (MGMS). The brown diamonds/contours represent the resolved MGMS (rMGMS) of PGC 34107. Its corresponding total MGMS covering a same region of CO(1-0) and stellar mass is shown in a red star. For reference (i.e. not a fit to the data) the black line shows a constant $\rm \log f_{H_2}$=-1, where $\rm \log f_{H_2}$=$\rm \log \Sigma_{gas}$ -$\rm \log \Sigma_*$. Other markers are same as shown in Panel A. The inclination correction is considered in both K-S law and MGMS.}
  \label{fig12}
\end{figure}

Panel A of Figure~\ref{fig12} shows the relationship between $\rm \Sigma_{gas}$ and $\rm \Sigma_{SFR}$, i.e., the Kennicutt-Schmidt (K-S) law, within the same region of ionized and molecular gas (i.e., the solid black polygon in Figure~\ref{fig10}), considering the correction of the inclination simultaneously (see section~\ref{sect2.2: Full Spectral Fitting}). Here we have already considered the contribution from the $\rm \Sigma_{HI}$, which is derived from the gas-phase metallicity by utilizing the empirical formula given by \cite{Schruba+18}. The final $\rm \Sigma_{gas}$ of PGC 34107 is about 2.52$\pm$0.14 $\rm M_\odot pc^{-2}$, listed in Table~\ref{tab1}. As shown in Figure~\ref{fig12}, the K-S law for the local ETGs (orange squares, \citealt{Davis+14}), spirals (green triangles/contours, \citealt{Shi+18}), and the nuclear regions of U/LIRGs (indigo squares, \citealt{Shi+18}) is displayed, and the K-S law for the high-redshift integrated main-sequence galaxies (red circles, \citealt{Shi+18}) are given as well. The best-fitting K-S law for \cite{Shi+18} sample is: 
\begin{equation}
\label{eq9}
{\rm \log \Sigma_{SFR} [M_\odot yr^{-1} kpc^{-2}] =1.41\times \log \Sigma_{gas}-3.97},
 \end{equation}
 Where $\rm \log \Sigma_{gas}$ is in the unit of $\rm M_\odot pc^{-2}$.
Considering the effect of the adopted IMF on comparison, here the $\rm \Sigma_{SFR}$ from different works has been calibrated based on a \cite{Salpeter+1955} IMF instead of a \cite{Kroupa+01} IMF. Simultaneously, we also ensure that here the $\rm \Sigma_{gas}$ from all works is calculated by adopting the same $\rm \alpha_{CO}$ as in this paper. But note that our results may be biased because the SFR and gas measurements are limited to a small region and the physical area adopted by literature makes a difference. In \cite{Davis+14} sample, most of ETGs are S0 galaxies, and distribute below the K-S law (black line) of \cite{Shi+18}, while PGC 34107 is consistent with it. It means that the northern star formation of PGC 34107 is enhanced compared to normal ETGs, and it basically follows the K-S law of star-forming galaxies. 
Within the same region of ionized and molecular gas, the SFE is about $\rm (1.74\pm 0.65) \times 10^{-9} yr^{-1}$, and it may take about 580 Myr to exhaust the gas if assuming a constant gas consumption without gas inflow. The depletion time of gas is much smaller than that ($\sim$2.5 Gyr) of normal ETGs (\citealt{Davis+15}). If we adopt the total molecular gas (see Section~\ref{sect3.2:CO(1-0)}: ($\rm 3.36\pm 1.1$) $\times 10^{8} M_\odot$) and the total SFR (see Section~\ref{sect2.2: Full Spectral Fitting}: $\rm 0.33\pm 0.01 M_\odot yr^{-1}$), the gas depletion time is about ($\rm1.02\pm 0.34$)$\times 10^9$ yr, which is consistent with that of \cite{Bigiel+08}, which is converted to be ($\rm 1.26\pm 0.5$) $\times 10^9$yr based on a Salpeter IMF. 

In addition, Panel B of Figure~\ref{fig12} shows the `molecular gas main sequence' (MGMS, \citealt{Lin+19}). The resolved MGMS (rMGMS) of PGC 34107 is shown in brown diamonds/contours and the corresponding total MGMS is shown as a red star, covering the same central region between molecular gas and stellar mass. A constant molecular gas fraction $\rm \log f_{H_2}$=-1 (defined as $\rm \log f_{H_2}$=$\rm \log \Sigma_{gas}$ -$\rm \log \Sigma_*$) is shown in a black line for reference. At a fixed $\rm \log \Sigma_*$,  the rMGMS and its integrated MGMS of PGC 34107 show higher $\rm \log \Sigma_{gas}$, which means there is a higher gas fraction in the center of PGC 34107, which is similar to the rMGMS of 12 local spirals of \cite{Shi+18}. We note that PGC 34107 lives in a low-density environment (\citealt{Xiao+16, Zhou+20}), so we propose a secular evolutionary scenario where the bar-induced molecular gas inflow accumulates the central gas reservoir and supplies the raw materials for the current star formation, then further promotes the star formation.

\section{Summary}
\label{sect5:Summary}
Based on 2D optical spectroscopy from CAHA and millimeter observation from NOEMA, we study a barred lenticular galaxy PGC 34107 with central star formation. The spatially resolved ionized and molecular gas provide us an opportunity to reveal the trigger and evolution of star formation in the center of S0s. Our main results are summarized as follows:

1. Based on the IFU spectroscopic observation, PGC 34107 with normal rotation disk shows younger age in the central region and higher surface mass density along the stellar bar. Most star formation is off-center and concentrated on the northern star-forming region, $\sim$380 pc away from the center.  

2. Revealed by NOEMA observation, the distribution of molecular gas CO(1-0) also mainly traces the northern star formation. The rotation of molecular gas is consistent with the rotation of stellar disk. Thanks to the high resolution of NOEMA, the blueshifted (redshifted) velocity component on the receding (approaching) side of galaxy is discovered along the stellar bar, which might suggest that there is a molecular gas inflow along the bar. The location of gas inflow ($\sim$ 300-550 pc) is connected to the peak of the off-center star formation ($\sim$380 pc), while the peak position might be associated with the inner Lindblad resonance (ILR), which can probably explain why the velocity of gas inflow gradually declines along the bar to the northern star-forming region. The plausible gas inflow provides an evidence that what kind of secular internal processes leads to the on-going star formation.

3. We also find the existence of the isotopic line of $^{13}$CO(1-0). Combining optical and millimeter observations, most $\rm H\alpha$, CO(1-0) and $^{13}$CO(1-0) emissions are found to be concentrated on the northern star-forming region. The integrated intensity line ratio $\Re_{10}=$ CO(1-0)/$^{13}$CO(1-0) for the whole galaxy is estimated about $7.52\pm0.01$, which is generally consistent with the ranges observed in spirals. The $\Re_{10}$ value is related to the intensity of the star formation.

4. PGC 34017 follows the local MZR and SFMS. Within the northern star-forming region covered by ionized and molecular gas simultaneously, its star formation also follows the K-S law, and the SFE is comparable to that of spirals. Based on rMGMS, PGC 34107 shows a higher gas fraction in the central region.
This might support a scenario that the gas inflow along the bar gradually accumulates the gas reservoir and provides the raw material of star formation, then triggers the star-forming activity and arranges the gas-phase metallicity, resulting in the active star formation similar to spirals.

\acknowledgments
The authors are very grateful to the anonymous referee for critical comments and instructive suggestions, which significantly strengthened the analyses in this work. This work is supported by the National Key Research and Development Program of China (No.2017YFA0402703), the National Natural Science Foundation of China (No. 11733002, 12121003, 12192220, and 12192222) and the science research grants from the China Manned Space Project with NO. CMS-CSST-2021-A05. This work is based on observations carried out with the IRAM Northern Extended Millimeter Array. IRAM is supported by INSU/CNRS (France), MPG (Germany) and IGN (Spain). In addition, we acknowledge the supports of the staffs from CAHA and NOEMA, especially Ana Lopez. LCH is supported by the National Science Foundation of China (11721303, 11991052) and the National Key R\&D Program of China (2016YFA0400702).
\software {KINEMETRY (\citealt{Krajnovic+06}), GILDAS (\citealt{Pety+05,Gildas+Team+13}), MPFIT (\citealt{Markwardt+09}), GALFIT (\citealt{Peng+02, Peng+10}), $\rm ^{3D}BAROLO$ (\citealt{Di+Teodoro+15}).}

\appendix
\section{Position-Velocity diagram}
\begin{figure}
  \begin{center}
  \includegraphics[width=0.45\textwidth, angle=0 ]{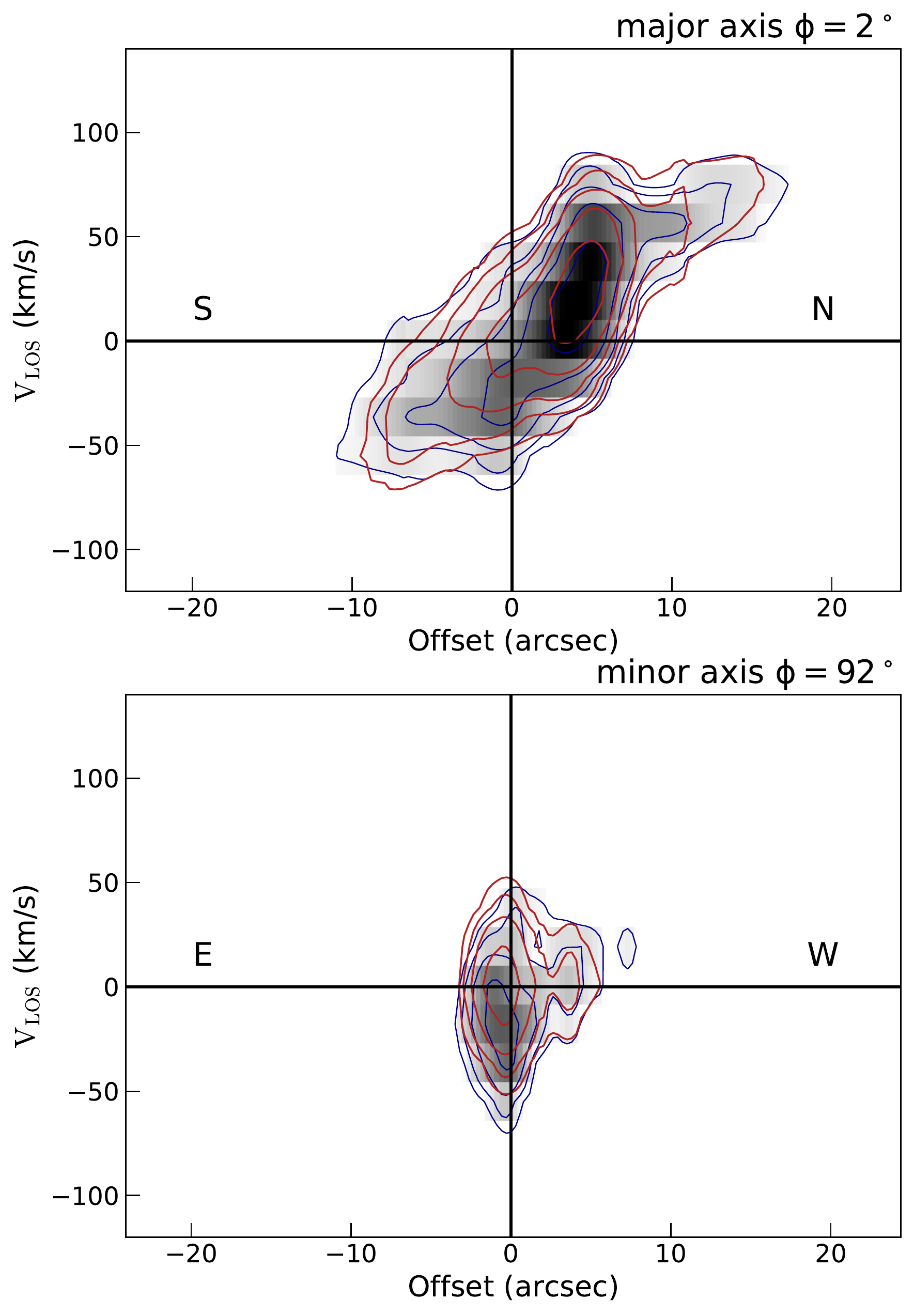}
  \end{center}
  \caption{The position-velocity diagrams (PVDs) for CO(1-0) emission line along the major axis (upper) at PA$\rm =2^\circ$ and minor axis (bottom) at $\rm 92^\circ$. The definition of PA is same as that derived by KINEMETRY. The blue contours represent the observed velocities. The best fit models from $\rm ^{3D}$BAROLO are shown as red contours, considering the corrections of both inclination and position angle. The levels of both contours are at $\rm 2^n \sigma$, from 2$\sigma$ to 64$\sigma$. The basic directions are also shown as S, N, E and W in the plots.} 
  \label{figA1}
\end{figure}

To model the rotation of molecular gas, we adopt $\rm ^{3D}$BAROLO software to perform a 3D tilted-ring fitting on the CO(1-0) emission line data cube (\citealt{Di+Teodoro+15}).  We set the position angle (PA), inclination ($i$), rotation velocity ($v_{\rm rot}$) and velocity dispersion ($v_{\rm disp}$) to be free after giving initial values (PA=0$^\circ$, $i$=80$^\circ$, $v_{\rm rot}$=100km/s, $v_{\rm disp}$=20km/s). The $\rm ^{3D}$BAROLO code fits a pure circular rotation model, so the non-circular motions, such as radial motion due to bar-induced gas inflow, can be reflect by comparing to the circular rotation model on the position-velocity diagram (PVDs, \citealt{Bureau+1999,Athanassoula+1999}). In Figure~\ref{figA1}, we show the results of PVDs from $\rm ^{3D}$BAROLO fitting. The upper panel is the PVD of CO(1-0) emission line along the major axis, which is closer to the pixels we extract, and the bottom is for minor axis.  The red contours are the best-fit results of circular motions by $\rm ^{3D}$BAROLO, while the blue contours and grey  mosaics are corresponding to the observed PVDs. Generally, the whole observed rotation follows a rotated orbit by comparing the shape of the model (red contours) with that of observation (blue contours). But specifically, the blue contours for the major axis in the north shift slightly lower than red contours, and the situation is inverse in the south, where blue ones tend to distribute above red ones, indicating that it is not a pure circular rotation along the major axis, which is also implied by the comparison of the minor axis between model and observation.

\bibliography{references}{}
\bibliographystyle{aasjournal}
\end{document}